\documentclass[preprint,amsmath,amssymb]{revtex4}

\usepackage{txfonts}    
\usepackage{color}     
\usepackage{graphicx}   
\usepackage{bm}

\newcommand{\e}{&\hspace{-.5em}}
\newcommand{\sub}[1]{\hspace{-.07em}\bm{#1}}
\newcommand{\subp}[1]{\hspace{-.07em}\bm{#1}_{\perp}}
\newcommand{\mr}[1]{\mathrm{#1} }
\newcommand{\para}{\parallel }

\newcommand{\odif}[2]{\ensuremath{\frac{d #1}{d #2 }}} 
\newcommand{\pdif}[2]{\ensuremath{\frac{\partial #1}{\partial #2}}}

\newcommand{\D}{\cdot}

\newcommand{\lbr}{\left ( }
\newcommand{\rbr}{\right ) }

\newcommand{\vpr}{\varv_{\parallel} }
\newcommand{\vpp}{\varv_{\perp} }
\newcommand{\vv}{\varv }
\newcommand{\bb }{\bm{b} }
\newcommand{\kp }{k_{\perp}^{2} }

\newcommand{\df }{\delta f }
\newcommand{\dfg }{\delta f^{\mathrm{(g)}} } 
\newcommand{\dfgk }{\dfg_{\mr{s}\sub{k}_{\perp}} }
\newcommand{\dsk }{\delta S_{\! \mr{s}\sub{k}_{\perp}} }
\newcommand{\tzf }{\mathcal{T}_{\mr{s}}^{\mr{(zf)}} }
\newcommand{\tzfi }{\mathcal{T}_{\mr{i}}^{\mr{(zf)}} }
\newcommand{\tzfe }{\mathcal{T}_{\mr{e}}^{\mr{(zf)}} }

\newcommand{\dphi }{\delta \phi }
\newcommand{\dphik }{\delta \phi_{\sub{k}_{\perp}} }
\newcommand{\dpsi }{\delta \psi }
\newcommand{\dpsik }{\delta \psi_{\sub{k}_{\perp} }}

\begin{document}


\title{Nonlinear entropy transfer via zonal flows in gyrokinetic plasma turbulence}
%
\author{Motoki Nakata$^\#$}
\affiliation{The Graduate University for Advanced Studies, Toki, Gifu 509-5292, Japan}
\thanks{$\#\,$Present affiliation: Faculty of Arts and Sciences, Komazawa University, Tokyo 154-8525, Japan \\ \textbf{This is the author accepted manuscript (AAM/AM), published in Physics of Plasmas 19, 022303 (2012)\\ doi: 10.1063/1.3675855}} 
\author{Tomo-Hiko Watanabe}
\author{Hideo Sugama}
\affiliation{National Institute for Fusion Science, Toki, Gifu 509-5292, Japan / \\
The Graduate University for Advanced Studies, Toki, Gifu 509-5292, Japan}
%
%
%

\begin{abstract}
Nonlinear entropy transfer processes in toroidal ion temperature gradient (ITG) and electron temperature gradient (ETG) driven turbulence 
are investigated based on the gyrokinetic entropy balance relations for zonal and non-zonal modes, which are coupled through 
the entropy transfer function regarded as a kinetic extension of the zonal-flow production due to the Reynolds stress. 
Spectral analyses of the ``triad'' entropy transfer function introduced in this study reveal not only the nonlinear interactions 
among the zonal and non-zonal modes, but also their effects on the turbulent transport level. 
Different types of the entropy transfer processes between the ITG and ETG turbulence are found: 
The entropy transfer from non-zonal to zonal modes is substantial in the saturation phase of the ITG instability, 
while, once the strong zonal flow is generated, the entropy transfer to the zonal modes becomes quite weak 
in the steady turbulence state. 
Instead, the zonal flows mediate the entropy transfer from non-zonal modes with low radial-wavenumbers (with contribution to the heat flux) 
to the other non-zonal modes with higher radial-wavenumbers (but with less contribution to the heat flux) through the triad interaction. 
The successive entropy transfer processes to the higher radial-wavenumber modes are associated with transport regulation in the steady turbulence state.
In contrast, in both the instability-saturation and steady phases of the ETG turbulence,
the entropy transfer processes among low-wavenumber non-zonal modes are dominant rather than the transfer via zonal modes. 
\end{abstract}

\maketitle
\section{INTRODUCTION}
Plasma turbulence driven by drift wave instabilities is a key issue for understanding anomalous transport of
particle, momentum, and heat observed in magnetically confined plasmas. 
Ion temperature gradient (ITG) and electron temperature gradient (ETG) driven instabilities
are considered as main causes of the micro-scale turbulence with the spatial scale of the ion and electron gyroradii, respectively.
Various flow structures, i.e., fine-scale turbulent vortices, axisymmetric zonal flows, and radially elongated streamers,
are generated through complicated nonlinear interactions in the plasma turbulence.

One of the most important findings for drift-wave turbulence, 
especially the ITG-driven turbulence, is effective regulation of the turbulent transport by self-generated zonal flows\cite{1,2,3,4}. 
Zonal flows are nonlinearly generated through the Reynolds stress resulting from turbulent flows in the drift-wave turbulence, 
where the radial scale-length of the zonal-flow shear is associated with a typical wavenumber range of the turbulence. 
Existence of ion-scale zonal flows has been experimentally revealed by a direct measurement of electrostatic potential in laboratory 
experiments\cite{5}. 
Tremendous efforts have been devoted so far to the study of the physical mechanisms of the zonal-flow generation, 
the nonlinear saturation of the linear instability growth, and the transport suppression. 
(A review of earlier works on the zonal-flow generation is given in Ref. 6.)

In the ETG turbulence, although the response kernel of the zonal flows has a larger residual amplitude, 
the source of the zonal-flow potential is shielded by the adiabatic responses of ions 
resulting from the perpendicular gyromotion with larger thermal gyroradii\cite{7,8}. 
Thus, the zonal-flow generation in the ETG turbulence is relatively weaker than that in the ITG turbulence 
so that the ETG turbulence involves not only zonal flows, but also various vortex structures, of which the appearance strongly depends 
on geometrical and plasma parameters\cite{8,9,10,11,12}.

From the aspect of regulating the turbulent transport in future burning plasmas,
it is worthwhile to understand fundamental physics behind the formation of vortex and zonal-flow structures and 
their nonlinear interactions as well as the related transport properties. 

In neutral fluid turbulence researches, the nonlinear interactions among turbulent flows/vortices has often been investigated with the energy transfer 
function, which appears in the energy balance equation derived from the Navier-Stokes equation\cite{13,14,15,16}.
Both the analytical and numerical investigations of the energy transfer function have contributed to understandings of the formation 
of steady turbulence spectra and to the construction of statistical closure model. 
In particular, the spectral analysis of the energy transfer for fully-developed isotropic turbulence 
has successfully revealed the local energy cascade processes in the inertial sub-range which is affected by neither forcing nor 
viscous dissipation, where the energy transfer function is integrated over the azimuthal angle on wavenumber-space (this is so called shell-average) 
so as to depend on the only wavenumber magnitude\cite{16}. 

On the other hand, flow structures in magnetized plasma turbulence are, in general, different from those 
in neutral fluid turbulence due to the strong anisotropic nature. 
Also, dynamics of turbulent vortices and zonal flows, and the related transport processes in high temperature 
collisionless (or weakly collisional) plasmas involves a lot of kinetic processes, i.e., the Landau damping, 
the finite gyroradius effect, the particle drift, and the magnetic trapping. 
However, the governing kinetic equation involves a quadratic nonlinearity so that the nonlinear interactions 
are characterized by triad interactions, which are similar to those in the Navier-Stokes equation. 

For the nonlinear interactions in the plasma kinetic turbulence, it is necessary to examine the entropy transfer function 
which should appear in the entropy balance equation\cite{17,18,19,20,21} describing explicitly the balance relation among 
the microscopic fluctuations of the distribution function, the turbulent transport flux, and the collisional dissipation.  
The entropy balance relations for non-zonal and zonal modes have a particular importance, 
because the entropy transfer function coupling these equations is regarded as a kinetic extension of zonal-flow energy production 
due to the Reynolds stress\cite{19}. 
The nonlinear energy (not entropy) transfer 
between zonal flows and the drift-wave turbulence has been discussed so far 
based on the theoretical analysis with a fluid model\cite{22,23} 
and on the bi-spectrum analysis for the experimental data of density (or potential) fluctuation\cite{24,25,26}. 
However, the turbulence diagnostics based on gyrokinetic entropy balance relations presented here can provide ones 
with more systematic method to quantify the effect of the nonlinear interaction on turbulent transport processes. 

Recently, some earlier works have discussed the entropy transfer processes in the drift-wave turbulence.  
Waltz \textit{et al}. have examined the spectral structures of the entropy transfer function 
for the ITG-TEM turbulence based on the gyrokinetic entropy balance, 
and have found that the entropy transfer to zonal modes is rather weak in the steady state\cite{27}. 
In addition, several works have revealed the local entropy transfer (or cascade) processes in 
sub-$\rho_{\mr{ti}}$ scale turbulence with slab\cite{28,29} and toroidal geometries\cite{30}, 
where the shell-averaged transfer function are used by assuming isotropic interactions.  
However, the role of zonal flows in the nonlinear entropy transfer processes in the steady turbulence state, 
which is crucial for determining the resultant transport level, has not been fully clarified yet. 
Moreover, it has not been revealed whether the entropy transfer processes are different 
between the saturation phase of the instability growth and the steady phase of turbulence. 

In order to address these issues, the non-averaged transfer function, which is a function of the wavenumber vectors, 
must be used so as to retain general anisotropic triad interactions. 
It is noted that the application of the shell-average obscures the nonlinear triad interactions with zonal modes. 
Moreover, for the quantitative evaluation of the role of zonal flows in entropy transfer, 
the detailed (or individual) entropy transfer process among triad modes should be investigated 
by means of the ``triad'' entropy transfer function [see Eq. (17)] and the detailed balance relation [see. Eq. (20)], 
beyond the analysis of gross entropy transfer function, which is summed over modes for one and/or two wavenumbers consisting a triad. 

In this paper, the nonlinear entropy transfer processes among non-zonal and zonal modes in toroidal ITG and ETG turbulence 
have been investigated by means of five-dimensional nonlinear gyrokinetic Vlasov simulations 
with the entropy balance and transfer diagnostics. 
In addition to examine gross entropy balance relations for non-zonal and zonal modes including the transfer term, 
spectral structures of the triad entropy transfer function are investigated in detail. 
The results presented here provide ones with a new understanding of the role of zonal flows in the entropy transfer processes, 
which is closely associated with the saturation of instability growth and transport suppression in the steady state. 

The remainder of this paper is organized as follows. 
A theoretical model used in the present study is described in Sec. II. 
Nonlinear gyrokinetic simulation results of the toroidal ITG and ETG turbulence are presented in Sec. III. 
Then, in addition to the entropy balance relations for non-zonal and zonal modes, wavenumber spectra of turbulent fluctuations 
and the heat flux are compared. 
In Sec. IV, differences between entropy transfer processes in the ITG and ETG turbulence are discussed through 
the comparisons of the wavenumber spectra of the triad entropy transfer function. 
Finally, concluding remarks are given in Sec. V. 

\section{THEORETICAL MODEL AND LINEAR STABILITY ANALYSIS}
Numerical simulations of the toroidal ITG and ETG turbulence presented here are carried out by using the GKV code\cite{2} 
based on the electrostatic gyrokinetic model. 
The electrostatic gyrokinetic equation for the perturbed distribution function $\dfgk$ written in the $\bm{k}_{\perp}$-space 
is given by 
%
\begin{multline} 
\left [\pdif{}{t}+\vpr \bb \! \D \nabla +i\omega_{\mr{Ds}} -\frac{\mu}{m_{\mr{s}}} \bb \! \D \nabla B \pdif{}{\vpr} \right ] 
\dfgk \\ 
\ \ \ \ \ \ \ \  - \frac{c}{B} \! \sum_{\Delta} 
\bb \D \lbr \bm{k}^{\prime}_{\perp}\times \bm{k}^{\prime \prime}_{\perp}
 \rbr \dpsi_{\sub{k}^{\prime}_{\perp}} \dfg_{\mr{s}\sub{k}^{\prime \prime}_{\perp}}\\
= F_{\mr{Ms}} \lbr i\omega_{\ast T\mr{s}} - i\omega_{\mr{Ds}} - \vpr \bb \D \nabla \rbr \frac{e_{\mr{s}}\dpsik}{T_{\mr{s}}}
+ \mathcal{C}_{\mr{s}} \! \left [ \dfgk \right ]\ ,
\end{multline}
%
where $\omega_{\mr{Ds}} \! \equiv \! (c/e_{\mr{s}}B)\bm{k}_{\perp} \! \D \bb \times 
(\mu \nabla B + m_{\mr{s}}\vpr^{2}\bb \! \D \! \nabla \bb )$ and 
$\omega_{\ast T\mr{s}} \! \equiv \! (cT_{\mr{s}}/e_{\mr{s}}B)\{1 \! +\! \eta_{\mr{s}}[(m_{\mr{s}}\vpr^{2} + \! 2\mu B)
/2T_{\mr{s}}-3/2]\} \bm{k}_{\perp} \! \D \bb \times \nabla \ln n_{\mr{s}}$ 
with $\eta_{\mr{s}} \! = \! | \nabla \! \ln T_{\mr{s}}| / |\nabla \! \ln n_{\mr{s}}|$ 
[the subscript ``$\mr{s}$'' denotes the particle species for ions ($\mr{s} \! = \! \mr{i}$) or electrons ($\mr{s} \! = \! \mr{e}$)]. 
Here, $\bm{b}$, $B$, $c$, $m_{\mr{s}}$, $n_{\mr{s}}$, $e_{\mr{s}}$, $T_{\mr{s}}$ and $\dpsik$ are 
the unit vector parallel to the magnetic field, 
the magnetic field strength, the speed of right, the particle mass, the particle number density, the electric charge, 
the equilibrium temperature 
and the electrostatic potential fluctuation averaged over the gyrophase, respectively. 
The symbol $\sum_{\bm{\Delta}}$ appearing in the nonlinear term of Eq. (1) stands for the summation over Fourier modes which satisfy
the triad-interaction condition, i.e., $\bm{k}_{\perp} \! = \! \bm{k}_{\perp}^{\prime}+\bm{k}_{\perp}^{\prime \prime}$. 
The parallel velocity $\vpr$ and the magnetic moment $\mu$ are used as the velocity-space 
coordinates, where $\mu$ is defined by $\mu \! \equiv \! m_{\mr{s}}\vpp^{2}/2B$ with the perpendicular velocity $\vpp$. 
The equilibrium part of the distribution function is given by the local Maxwellian distribution, i.e., 
$F_{\mr{Ms}} \! = \! n_{\mr{s}}(m_{\mr{s}}/2\pi T_{\mr{s}})^{3/2}\exp[-(m_{\mr{s}}\vpr^{2}+2\mu B)/2T_{\mr{s}}]$. 
A weak collisional effect is introduced in terms of a reduced collision operator given by 
%
\begin{equation} 
\mathcal{C}_{\mr{s}} = \nu_{\mr{s}} \left [  \frac{1}{\vpp}\pdif{}{\vpp} \lbr \vpp \pdif{}{\vpp} + 
\frac{\vpp^{2}}{\vv_{\mr{ts}}^{2}} \rbr + \pdif{}{\vpr} \lbr \pdif{}{\vpr} + \frac{\vpr}{\vv_{\mr{ts}}^{2}}\rbr \right ]\ , 
\end{equation}
%
where $\nu_{\mr{s}}$ and $\vv_{\mr{ts}} \! \equiv \! (T_{\mr{s}}/m_{\mr{s}})^{1/2} $ are the collision frequency and the thermal speed, 
respectively. 
The collision operator acting on $\dfgk$ smooths out the fine-scale fluctuations in the velocity-space. 
The usage of the above reduced collision operator, which breaks conservation properties for the momentum and the energy, 
may influence fine-scale velocity-space structures of the perturbed distribution function. 
However, the simplified collision term does not affect dynamics in long- and middle- wavelength regime, 
which are relevant to the heat transport and the entropy transfer processes in wavenumber-space discussed below,  
as long as the collision frequency is small enough in comparison with the thermal transit frequency\cite{21}, 
i.e., $\nu_{\mr{s}}L_{n_{0}}/\vv_{\mr{ts}} \! \ll \! 1$. 

A local toroidal flux-tube system in a large-aspect-ratio tokamak with the concentric circular flux surfaces is considered here\cite{31}. 
The magnetic field is given by $\bm{B} \! = \! B[\bm{\mr{e}}_{\zeta}+(r/R_{0})\bm{\mr{e}}_{\theta}]$ with neglecting 
$\mathcal{O}[(r/R_{0})^{2}]$ terms, where $r$, $R_{0}$, $q$, $\bm{\mr{e}}_{\zeta}$ and $\bm{\mr{e}}_{\theta}$ denote 
the minor and the major radii, the safety factor, the unit vectors in the toroidal($\zeta$) and poloidal($\theta$) 
directions, respectively.  
The field-aligned coordinates ($x,y,z$) are, then, introduced by $x \! = \! r-r_{0}$, 
$y \! = \! (r_{0}/q_{0})[q\theta - \zeta]$, $z \! = \! \theta$ with $q\! = \! q_{0}[1+\hat{s}(r-r_{0})/r_{0}]$, 
where $r_{0}$ and $\hat{s}$ represent the radial position of the flux-tube center and the magnetic shear parameter assumed 
to be constant, respectively. 
Use of the field-aligned coordinates enables us to impose the periodic boundary condition in the $x$- and $y$-directions 
so that the nonlinear $\bm{E} \! \times \! \bm{B}$ convection term [the last term in the left-hand side of Eq. (1)] 
is calculated numerically by means of the Fourier spectral method, 
where the perpendicular wavenumber vector $\bm{k}_{\perp}$ and the squared norm $\kp$ are written as 
$\bm{k}_{\perp} \! = \! k_{x} \nabla x + k_{y}\nabla y$ and $\kp \! = \! (k_{x}+\hat{s}zk_{y})^{2}+k_{y}^{2}$, respectively. 
(Note here that $\bb \D \nabla x \! = \! 0$ and $\bb \D \nabla y \! = \!0$, but $\nabla x \D \nabla y \! \neq \! 0$.) 
In this coordinate system, the magnetic field strength $B$ is reduced to $B \! = \! B_{0}(1-\epsilon \cos z)$ 
with $\epsilon \! \equiv \! r_{0}/R_{0}$, and then the operators 
$\bb \! \D \nabla$, $\omega_{\mr{Ds}}$ and $\omega_{\ast T\mr{s}}$ in Eq. (1) are written as 
%
\begin{eqnarray} 
\bb \! \D \nabla \ \e = \ \e \frac{1}{q_{0}R_{0}}\pdif{}{z}\ , \\
\omega_{\mr{Ds}} \ \e = \ \e -\sigma_{\mr{s}} \frac{\vpr^{2} \! + \!\mu B}{\vv_{\mr{ts}}R_{0}} 
    \left [ k_{x}\rho_{\mr{ts}} \sin z  + k_{y}\rho_{\mr{ts}} \lbr \cos z +\hat{s} z \sin z \rbr \right ], \\
\omega_{\ast T\mr{s}} \ \e = \ \e -\sigma_{\mr{s}}\frac{\vv_{\mr{ts}}}{L_{n_{\mr{s}}}}
    \left [ 1 + \eta_{\mr{s}} \lbr \frac{m_{\mr{s}}\vpr^{2} +  2\mu B}{2T_{\mr{s}}} - \frac{3}{2} \rbr \right ] k_{y}\rho_{\mr{ts}}\ ,
\end{eqnarray}
%
in the low-$\beta$ limit, respectively, where $\sigma_{\mr{s}} \! = \{ 1$ (for $\mr{s} \! = \! \mr{i}$), $-1$ (for $\mr{s} \! = \! \mr{e}$)\} 
and the thermal gyroradius is denoted by $\rho_{\mr{ts}} \! = \! (\vv_{\mr{ts}}/\Omega_{\mr{s}})|_{B=B_{0}}$ 
with the gyrofrequency $\Omega_{\mr{s}} \! = \! m_{\mr{s}}c/|e_{\mr{s}}|B$. 
The gradient scale-length of the density profile is represented by $L_{n_{\mr{s}}} \! \equiv \! |\nabla \! \ln n_{\mr{s}}|^{-1}$. 

The potential fluctuation evaluated at the particle position, $\dphik$, is related to the gyrophase-averaged one, 
i.e., $\dpsik \! = \! J_{0\mr{s}}\dphik$, and is determined by the Poisson equation written in the wavenumber-space as follows, 
%
\begin{eqnarray}
k_{\perp}^{2}\lambda_{\mr{De}}^{2}n_{0} \! \frac{e\dphik}{T_{\mr{e}}} & = & \left [ \int \! \! 
d\bm{\vv} J_{0\mr{i}}\dfg_{\mr{i}\sub{k}_{\perp}} \!
-  n_{0} \! \frac{e\dphik}{T_{\mr{i}}}\lbr 1-\Gamma_{0\mr{i}} \rbr \right ]  \nonumber \\
& - & \left [ \int \! \! d\bm{\vv} J_{0\mr{e}}\dfg_{\mr{e}\sub{k}_{\perp}} \!
+ n_{0} \! \frac{e\dphik}{T_{\mr{e}}}\lbr 1-\Gamma_{0\mr{e}} \rbr \right ],
\end{eqnarray}   
%
where $|e_{\mr{i}}| \! = \! |e_{\mr{e}}| \! = \! e$ and $n_{\mr{i}}\! =\! n_\mr{{e}}\! =\! n_{0}$ are assumed. 
The electron Debye-length is denoted by $\lambda_{\mr{De}} \! \equiv \! (T_{\mr{e}}/4\pi n_{0}e^{2})^{1/2}$. 
The first and the second groups of terms on the right hand side of Eq. (6) indicate the ion and electron density 
fluctuations represented with the gyrocenter distribution function and the electrostatic potential, respectively. 
Here, $J_{0\mr{s}}$ and $\Gamma_{0\mr{s}}$ are defined by 
$J_{0\mr{s}} \! \equiv \! J_{0}(k_{\perp}\vpp/\Omega_{\mr{s}})$ and 
$\Gamma_{0\mr{s}} \! \equiv \! I_{0}(b_{\mr{s}})\exp(-b_{\mr{s}})$ with the zeroth-order Bessel and modified Bessel functions for 
$b_{\mr{s}} \! \equiv \! \kp \vv_{\mr{ts}}^{2}/\Omega_{\mr{s}}^{2}$, respectively. 
For the ITG turbulence with $k_{\perp}\rho_{\mr{te}} \! \ll \! 1$, 
the adiabatic electron response is assumed (except for zonal modes) so that Eq. (6) is reduced to\cite{32} 
%
\begin{equation}
\int \! \! 
d\bm{\vv} J_{0\mr{i}}\dfg_{\mr{i}\sub{k}_{\perp}} \!
= n_{0} \Lambda_{\mr{i}\sub{k}_{\perp}} \! \frac{e\dphik}{T_{\mr{i}}} - n_{0} \left \langle \frac{e\dphik}{T_\mr{e}{}} \right \rangle 
\delta_{k_{y},0}\ \ \text{( for ITG )},
\end{equation}   
%
where $\Lambda_{\mr{i}\sub{k}_{\perp}} \! \equiv \! 1+T_{\mr{i}}/T_{\mr{e}}-\Gamma_{0\mr{i}}$ and $\delta_{m,n}$ is the Kronecker delta. 
The angular brackets $\langle \cdots \rangle$ stand for the field line average, i.e., 
$\langle A_{\sub{k}_{\perp}} \rangle \! \equiv \! \int \! dz A_{\sub{k}_{\perp}}B^{-1} / \! \int \! dz B^{-1}$. 
Thus, $\langle A_{\subp{k}} \rangle \delta_{k_{y},\,0}$ is equivalent to the flux surface average of $A$. 
For the ETG turbulence with $k_{\perp}\rho_{\mr{ti}} \! \gg \! 1$, 
the ion response to the potential fluctuation is reduced to the adiabatic one because of $J_{0\mr{i}} \! \ll \! 1$ and 
$\Gamma_{0\mr{i}} \! \ll \! 1$. 
Then, Eq. (6) is rewritten as 
%
\begin{equation}
\int \! \! 
d\bm{\vv} J_{0\mr{e}}\dfg_{\mr{e}\sub{k}_{\perp}} \!
= -n_{0} \Lambda_{\mr{e}\sub{k}_{\perp}} \! \frac{e\dphik}{T_{\mr{e}}}\ \ \text{( for ETG )}.
\end{equation}
%
Here, $\Lambda_{\mr{e}\sub{k}_{\perp}} \! \equiv \! 1+T_{\mr{e}}/T_{\mr{i}}-\Gamma_{0\mr{e}}$, and the finite Debye length effect 
$\kp\lambda_{\mr{De}}^{2}$ is ignored. 

Using the closed set of equations described above, one obtains a balance equation with respect to 
the entropy variable $\dsk$ defined as a functional of the perturbed gyrocenter distribution function $\dfgk$. 
The velocity space integral and the field line average of Eq. (1) multiplied by $\df_{\mr{s}\subp{k}}^{\mr{(g)\ast}}/F_{\mr{Ms}}$ lead to 
%
\begin{equation}
\pdif{}{t}\lbr \dsk \! + W_{\mr{s}\subp{k}} \rbr = L_{T_{\mr{s}}}^{-1}Q_{\mr{s}\subp{k}} \! 
+ \mathcal{T}_{\mr{s}\subp{k}} \! +D_{\mr{s}\subp{k}}\ ,
\end {equation}
%
where
%
\begin{eqnarray}
\dsk \e \equiv \e \  \left \langle \int \! \! d\bm{\vv} \, \frac{|\dfgk|^{2}}{2F_{\mr{Ms}}} \right \rangle\ , \\
Q_{\mr{s}\subp{k}} 
     \e \equiv \e \ \mr{Re} \left \langle i \vv_{\mr{ts}} \! \! \int \! \! d\bm{\vv} \, 
                    \df_{\mr{s}\subp{k}}^{\mathrm{(g)}} 
                 \lbr \frac{m_{\mr{s}}\vpr^{2}+2\mu B}{2T_{\mr{s}}} \rbr k_{y}\rho_{\mr{ts}} 
                 \frac{e\dpsik^{\ast}}{T_{\mr{s}}} \right \rangle,\\
D_{\mr{s}\subp{k}} 
     \e \equiv \e \ \mr{Re} \left \langle \int \! \! d\bm{\vv} \, \mathcal{C}_{\mr{s}} \! \left [ \dfgk \right ]
             \frac{h_{\mr{s}\subp{k}}^{\ast}}{F_{\mr{Ms}}} \right \rangle \ ,
\end {eqnarray}
%
denote the entropy variable, the turbulent heat flux, and the collisional dissipation, respectively. 
(Note that no particle flux is driven by the turbulence with the adiabatic response of background particles.) 
The gradient scale-length of the temperature profile is represented by $L_{T_{\mr{s}}} \! \equiv \! |\nabla \! \ln T_{\mr{s}}|^{-1}$. 
The non-adiabatic part of the perturbed gyrocenter distribution function, $h_{\mr{s}\subp{k}}$, is defined by 
%
\begin{equation}
\dfgk = -\frac{e_{\mr{s}}\dpsik}{T_{\mr{s}}}F_{\mr{Ms}} + h_{\mr{s}\subp{k}}\ .
\end {equation}
%

One can easily show the self-adjointness for the model collision operator used here [see Eq. (2)] by taking a partial integral 
so that the velocity-moment of $\mathcal{C}_{\mr{s}}[\mathcal{G}]\mathcal{G}/F_{\mr{Ms}}$ shows the negative-definite property 
for an arbitrary function $\mathcal{G}(\vpr,\, \mu)$. 
In the present simulation model, the collision term acting on the perturbed gyrocenter distribution function 
$\mathcal{C}_{\mr{s}}[\dfgk]$ is used for simplicity, 
instead of its non-adiabatic part $h_{\mr{s}\subp{k}}$ so that the collisional dissipation in Eq. (12) does not have 
the exact negative-definite form. 
Even with this simplification, we found that the collisional dissipation still keeps a negative value, 
as the contribution of the adiabatic part $-(e_{\mr{s}}\dpsik/T_{\mr{s}})F_{\mr{Ms}}$ is small enough. 

By using Eqs. (7) and (8), the potential energy $W_{\mr{s}\subp{k}}$ is given by 
%
\begin{eqnarray}
W_{\mr{i}\subp{k}} \e = \e \ \frac{n_{0}}{2} \left \langle  \lbr \Lambda_{\mr{i}\subp{k}} \!  - \frac{T_{\mr{i}}}{T_{\mr{e}}} 
\delta_{k_{y},0} \rbr \left | \frac{e\dphik}{T_{\mr{i}}} \right |^{2} \right \rangle\ , \\
W_{\mr{e}\subp{k}} \e = \e \ \frac{n_{0}}{2} \left \langle  \Lambda_{\mr{e}\subp{k}} \left | \frac{e\dphik}{T_{\mr{e}}} \right |^{2} 
\right \rangle\ ,
\end {eqnarray}
%
for ions and electrons, respectively. 

The second term in the right hand side of Eq. (9) represents the nonlinear entropy transfer in the wavenumber space. 
The definition of the entropy transfer function $\mathcal{T}_{\mr{s}\subp{k}}$ is given by 
%
\begin{eqnarray}
\mathcal{T}_{\mr{s}\subp{k}}
\e \ = \e \ \sum_{\subp{p}} \! \sum_{\subp{q}}\delta_{\subp{k}\! +\subp{p}\! +\,\subp{q},\,0}\, 
\mathcal{J}_{\mr{s}}\left [ \bm{k}_{\perp}| \bm{p}_{\perp}, \bm{q}_{\perp} \right ]\ ,  \\
\mathcal{J}_{\mr{s}}\left [ \bm{k}_{\perp}| \bm{p}_{\perp}, \bm{q}_{\perp} \right ]
\e \ \equiv \e \ 
\left \langle \frac{c}{B} \bb \D \lbr \bm{p}_{\perp}\times \bm{q}_{\perp} \rbr \! \int \! \! d\bm{\vv} \, \frac{1}{2F_{\mr{Ms}}} 
\nonumber \right. \\ 
\times \e \e  \! \! \! \! \mr{Re} 
\left. \left [ \dpsi_{\subp{p}} h_{\mr{s}\subp{q}} h_{\mr{s}\subp{k}} \! - \dpsi_{\subp{q}} h_{\mr{s}\subp{p}} h_{\mr{s}\subp{k}} \right ] 
\! \frac{}{}\right \rangle ,
\end{eqnarray}
%
where the notation with $\bm{k}_{\perp}^{\prime}$ and $\bm{k}_{\perp}^{\prime \prime}$ shown in Eq. (1) is replaced here by 
$-\bm{p}_{\perp}$ and $-\bm{q}_{\perp}$, respectively, in order to represent symmetrically the triad-interaction condition 
for three wavenumber vectors, i.e., $\bm{k}_{\perp} \! +\bm{p}_{\perp} \! +\bm{q}_{\perp} \! = \! 0$. 
In Eq. (16), $\mathcal{J}_{\mr{s}}[ \bm{k}_{\perp}| \bm{p}_{\perp}, \bm{q}_{\perp} ]$ is summed over $\bm{p}_{\perp}$ and 
$\bm{q}_{\perp}$. 
For convenience, we call the function $\mathcal{J}_{\mr{s}}[ \bm{k}_{\perp}| \bm{p}_{\perp}, \bm{q}_{\perp} ]$ 
the ``triad (entropy) transfer function'', hereafter. 
It should be noted that the triad transfer function possesses the following symmetry properties, 
%
\begin{eqnarray}
\mathcal{J}_{\mr{s}}\left [ \bm{k}_{\perp}| \bm{p}_{\perp}, \bm{q}_{\perp} \right ] 
\e \ = \e \ \mathcal{J}_{\mr{s}}\left [ \bm{k}_{\perp}| \bm{q}_{\perp}, \bm{p}_{\perp} \right ]\ ,\\
\mathcal{J}_{\mr{s}}\left [ \bm{k}_{\perp}| \bm{p}_{\perp}, \bm{q}_{\perp} \right ]
\e \ = \e \ \mathcal{J}_{\mr{s}}\left [ -\bm{k}_{\perp}| - \! \bm{p}_{\perp}, -\bm{q}_{\perp} \right ]\ .
\end {eqnarray}
%
Furthermore, one obtains straightforwardly the ``detailed balance relation'' for the triad-interactions, 
%
\begin{equation}
 \mathcal{J}_{\mr{s}}\left [ \bm{k}_{\perp}| \bm{p}_{\perp}, \bm{q}_{\perp} \right ] 
+\mathcal{J}_{\mr{s}}\left [ \bm{p}_{\perp}| \bm{q}_{\perp}, \bm{k}_{\perp} \right ] 
+\mathcal{J}_{\mr{s}}\left [ \bm{q}_{\perp}| \bm{k}_{\perp}, \bm{p}_{\perp} \right ] 
=0\ .
\end {equation}
%
Positive values of the triad transfer function $\mathcal{J}_{\mr{s}}[ \bm{k}_{\perp}| \bm{p}_{\perp}, \bm{q}_{\perp} ]$ mean that 
the entropy is transferred from two modes with $\subp{p}$ and $\subp{q}$ toward the mode with $\subp{k} (= \! -\subp{p}\! -\subp{q})$ 
and that the possible combinations of the signs of 
($\mathcal{J}_{\mr{s}} [ \bm{k}_{\perp}| \bm{p}_{\perp}, \bm{q}_{\perp}  ]$,   
$\mathcal{J}_{\mr{s}} [ \bm{p}_{\perp}| \bm{q}_{\perp}, \bm{k}_{\perp} ]$,   
$\mathcal{J}_{\mr{s}} [ \bm{q}_{\perp}| \bm{k}_{\perp}, \bm{p}_{\perp} ]$) satisfying the detailed balance relation 
should be ($+,-,-$), ($+,+,-$) and ($+,-,+$). 
Negative values of $\mathcal{J}_{\mr{s}}[ \bm{k}_{\perp}| \bm{p}_{\perp}, \bm{q}_{\perp} ]$ 
indicate the entropy transfer in the opposite direction, 
then ($-,+,+$), ($-,-,+$) and ($-,+,-$) are the possible combinations of the signs of the triad transfer functions. 
The detailed balance relation Eq. (20) with the symmetric properties of Eqs. (18) and (19) is useful for the entropy transfer 
analysis of the toroidal ITG and ETG turbulence simulation results shown in Sec. IV, where the nonlinear interactions among 
turbulent fluctuations and zonal flows are quantified. 

The definition of the triad entropy transfer function Eq. (17) follows a method used for the ``energy transfer function'' 
in neutral fluid turbulence governed by the Navier-Stokes equation\cite{16}. 
As isotropic turbulence is considered, the energy transfer function is often integrated over the azimuthal angle on the wavenumber space 
(this is so called shell-average), and is reduced to a function depending on the only wavenumber magnitude. 
Then, the several analytic expressions are derived with statistical closure models 
such as the quasi-normal Markovian (QNM) model\cite{13,16} and the direct interaction approximation (DIA)\cite{14,15}. 
The spectral analysis of the energy transfer function for fully-developed turbulence has successfully revealed 
the local energy cascade processes in the inertial sub-range\cite{16}. 

On the other hand, the (quasi) two-dimensional magnetized plasma turbulence is often anisotropic 
due to the existence of zonal flows and streamers which are elongated flow structures in particular directions. 
For the sub-$\rho_{\mr{ti}}$ scale plasma turbulence, the local entropy transfer has been shown 
by means of the shell-averaged entropy transfer function\cite{28,29,30}.       
In order to discuss the entropy transfer among zonal and non-zonal modes in the ITG and ETG turbulence, 
the shell-averaged transfer function is not suitable, and the above triad transfer function [Eq. (17)] defined as a function of 
the wavenumber vectors should be used so as to retain general anisotropic triad interactions. 
(Note that the triad transfer function is not restricted to anisotropic interactions only, but involves isotropic ones 
when the shell-average is applied.) 

Also, the non-symmetrized triad transfer function, which does not hold the symmetry for the interchange of $\bm{p}_{\perp}$ 
and $\bm{q}_{\perp}$ [ Eqs. (18) ], has been used in earlier works. 
However, it should be pointed out that the magnitude of non-symmetrized one involves a spurious value resulting from 
the anti-symmetric part, which does not contribute to the gyrokinetic equation at all, 
even though its gross magnitude after the summation over $\bm{p}_{\perp}$ and $\bm{q}_{\perp}$ is same as that of the symmetrized one. 
(see also Appendix for discussions on the symmetrized transfer function.)  
Thus, the symmetrized triad transfer function presented here must be used for quantitative evaluation of the effect of 
the detailed (or individual) entropy transfer on turbulent transport level. 
%
\begin{figure}
\centering
\includegraphics[scale=1.0]{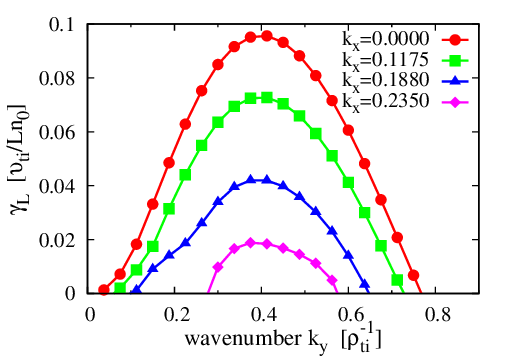}
\caption{(Color online) Wavenumber spectrum of the linear growth rate $\gamma_{\mr{L}}$ of the toroidal ITG instability. 
The spectrum for the toroidal ETG instability is the same as that in the ITG case except for 
the normalizations with $\vv_{\mr{te}}$ and $\rho_{\mr{te}}$. }
\end{figure}
%

The explicit description of the entropy balance relation for the non-zonal and zonal modes 
is useful for the following discussions concerning the entropy transfer processes among zonal flows and turbulence. 
By taking the integration of Eq. (9) over the non-zonal ($k_{y} \! \neq \! 0$) and zonal modes ($k_{y}\! =\! 0$) 
and by using the detailed balance relation of Eq. (20), one obtains 
%
\begin{eqnarray}
\odif{}{t} \lbr \delta S_{\! \mr{s}}^{\mr{(trb)}} + W_{\mr{s}}^{\mr{(trb)}} \rbr = 
L_{T_{\mr{s}}}^{-1}Q_{\mr{s}} -\tzf + D_{\mr{s}}^{\mr{(trb)}}\ , \\
\odif{}{t} \lbr \delta S_{\! \mr{s}}^{\mr{(zf)}} + W_{\mr{s}}^{\mr{(zf)}} \rbr = \tzf + D_{\mr{s}}^{\mr{(zf)}}\ ,
\end {eqnarray}
%
where the superscripts ``(trb)'' and ``(zf)'' represent the turbulence (or non-zonal) and zonal-flow components, respectively. 
(Here, the zonal modes are simply defined by $k_{y}\! =\! 0$ because Eq. (9) is already averaged over the field-line.)
The above entropy balance relations for non-zonal and zonal modes have been derived and discussed 
in detail by Sugama \textit{et al}.\cite{19}. 
Note here that zonal flows never contribute to the radial heat flux $Q_{\mr{s}\subp{k}}$ as seen from Eq. (22). 
The entropy transfer function integrated over the zonal modes, $\tzf$, represents the entropy transfer from turbulence to zonal flows 
so that the negative sign of $\tzf$ appears in Eq. (21).
As shown in Ref. 19, by using the simplest approximation for the non-adiabatic part of the ion gyrocenter distribution function, i.e., 
$h_{\mr{i}\subp{k}} \! \! \simeq \! n_{0}F_{\mr{Mi}}(1+\kp\rho_{\mr{ti}}^{2}/2)e\dphik /T_{\mr{i}}$, 
the entropy transfer function $\tzfi$ reduces to the energy production term, which is described by the product 
of the Reynolds stress due to the non-zonal turbulent flows and the zonal-flow shear. 
Thus, $\tzf$ is regarded as a kinetic extension of the zonal-flow energy production due to the Reynolds stress. 

As seen from Eq. (21), the turbulence part of the entropy $\delta S_{\! \mr{s}}^{\mr{(trb)}}$ is 
produced by the turbulent heat flux $L_{T_{\mr{s}}}^{-1}Q_{\mr{s}}$, partly dissipated by collisions $D_{\mr{s}}^{\mr{(trb)}}$, 
and partly transferred to the zonal flow components via the transfer term $\tzf$. 
When the turbulence reaches a statistically steady state, the balance relations of 
$\overline{\mathcal T}_{\! \mr{s}}^{\mr{(zf)}} \! \!  = \! -\overline{D}_{\mr{s}}^{\mr{(zf)}} \! \geqslant \! 0$ 
and $L_{T_{\mr{s}}}^{-1}\overline{Q}_{\mr{s}}-\overline{\mathcal T}_{\! \mr{s}}^{\mr{(zf)}}\! \! =\! -\overline{D}_{\mr{s}}^{\mr{(trb)}} 
\! \geqslant \! 0$ are realized separately, where the overline denotes the time-average in a steady phase. 
The entropy balance relations of Eqs. (21) and (22) provide us with not only the physical insight into the entropy transfer processes 
among zonal flows and turbulence, but also a good measure for the numerical accuracy of turbulence simulations.

The nonlinear gyrokinetic simulation results, as will be shown below, have been obtained by means of the GKV code, 
that is a gyrokinetic Vlasov solver with the flux-tube configuration applicable to both the tokamak and helical systems\cite{2}. 
The physical parameters of 
$\epsilon \! = \! 0.18$, $q_{0} \! = \! 1.5$, $\hat{s} \!= \! 0.4$, $R_{0}/L_{T_{\mr{s}}} \! = \! 6.92$, $\eta_{\mr{s}} \! = \! 2.0$ 
and $T_{\mr{i}} \! = \! T_{\mr{e}}$ are the same as the Cyclone-base case, 
except that the smaller values of the magnetic-shear parameter $\hat{s}$ and $\eta_{\mr{s}}$ 
are used here to make it easier to carry out the ETG turbulence simulation. 
The weak collisional effect is introduced with $\nu_{\mr{s}}L_{n_{0}}/\vv_{\mr{ts}} \! = \! 10^{-3}$. 
The number of Fourier modes in the perpendicular wavenumber space and the number of grids in the $z$-, $\vpr$-, and $\mu$-directions 
are set to be ($N_{k_{x}},N_{k_{y}},N_{z},N_{\vpr},N_{\mu}$)$=$($64,129,128,64,32$). 
The corresponding ranges of the phase-space coordinates are given as 
$0 \! \leqslant \! k_{x} \! \leqslant \! k_{x\mr{(max)}} \! =\! 1.5\rho_{\mr{ts}}^{-1}$, 
$-k_{y\mr{(max)}} \! \leqslant \! k_{y} \! \leqslant \! k_{y\mr{(max)}} \! =\! 2.4\rho_{\mr{ts}}^{-1}$,  
$-\pi \! \leqslant \! z \! \leqslant \! \pi$, 
$-5\vv_{\mr{ts}} \! \leqslant \! \vpr \! \leqslant \! 5\vv_{\mr{ts}}$, and 
$0 \! \leqslant \! \mu \! \leqslant \! 12.5 (m_{\mr{s}}\vv_{\mr{ts}}^{2}/B_{0})$, respectively. 
The modified periodic boundary condition is imposed in the $z$-direction\cite{31}. 
The convergence of transport level and entropy transfer has also been confirmed, 
as the range of $z$-direction is extended to $-3\pi \! \leqslant \! z \! \leqslant \! 3\pi$ with keeping the same resolution. 
The size of the perpendicular domain is $L_{x} \! \times \! L_{y}= \! 266 \rho_{\mr{ts}} \! \times \! 168\rho_{\mr{ts}}$, 
and the non-zero minimum absolute values of the wavenumber are $k_{x\mr{(min)}} \! = \! 0.0235\rho_{\mr{ts}}^{-1}$ and 
$k_{y\mr{(min)}} \! = \! 0.0375\rho_{\mr{ts}}^{-1}$. 
The time integration is carried out by means of the fourth-order Runge-Kutta-Gill method 
with $\Delta t \! =\! 0.025(L_{\mr{n_{0}}}/\vv_{\mr{ts}})$, 
where the timestep is chosen to resolve the parallel particle streaming, which is much faster than the perpendicular 
$\bm{E} \! \times \! \bm{B}$- and the magnetic drifts. 
The Courant number for the parallel streaming is given as 
$C_{\para} \! \equiv \! (\vpr \Delta t)/(q_{0}R_{0}\Delta z) \! \simeq \! 0.49$.
%

In the followings, the physical quantities are normalized as 
$x\! = \! x^{\prime}/\rho_{\mr{ts}},\ y \! = \! y^{\prime}/\rho_{\mr{ts}},\ \vpr \! = \! \vpr^{\prime}/\vv_{\mr{ts}},
\ \mu \! = \! \mu^{\prime}(B_{0}/m_{\mr{s}}\vv_{\mr{ts}}^{2}),
\ t \! = \! t^{\prime} (\vv_{\mr{ts}}/L_{n_{0}}),\ \nu_{\mr{s}}\! = \! \nu_{\mr{s}}^{\prime}(L_{n_{0}}/\vv_{\mr{ts}}),\ 
F_{\mr{Ms}}\! = \! F_{\mr{Ms}}^{\prime}(\vv_{\mr{ts}}^{3}/n_{\mr{0}}),\ 
\dfgk \! = \! \df_{\mr{s}\subp{k}}^{\mr{(g)}\prime}(\vv_{\mr{ts}}^{3}/
n_{\mr{0}})(L_{n_{0}}/\rho_{\mr{ts}})$ and $\dphik \! = \! \dphik^{\prime}(e/T_{\mr{s}})(L_{n_{0}}/\rho_{\mr{ts}})$, 
where the prime means a dimensional quantity. 

The wavenumber spectrum of the linear growth rate $\gamma_{\mr{L}}$ for the toroidal ITG instability is shown 
in Fig. 1, where the physical parameters shown above are used. 
One finds that the maximum growth rate of $\gamma_{\mr{L(max)}} \! =\! 9.56 \! \times \! 10^{-2}\vv_{\mr{ti}}/L_{n_{0}}$
is observed at ($k_{x} \! = \!0$, $k_{y}\! = \! 0.4125 \rho_{\mr{ti}}^{-1}$). 
The effect of the finite $k_{x}$ reduces the growth rate, then the toroidal ITG instability are completely stabilized for 
$k_{x} \! \geqslant \! 0.3\rho_{\mr{ti}}^{-1}$. 
As will be discussed below, the stabilizing effect in the higher $k_{x}$ region is important for the regulation of turbulent transport 
by zonal flows. 
Also, note that the spectrum of $\gamma_{\mr{L}}$ for the toroidal ETG instability is the same as that plotted in Fig. 1 
except for the normalizations with $\vv_{\mr{te}}$ and $\rho_{\mr{te}}$. 

\section{NONLINEAR SIMULATIONS}  
\subsection{Entropy balance relation}
The results of nonlinear gyrokinetic simulations for the toroidal ITG and ETG turbulence are shown and discussed in this section. 
Time evolutions of each term in the entropy balance relation of the turbulence part, Eq. (21), 
for the toroidal ITG and ETG turbulence are plotted in Figs. 2(a) and 2(b), respectively. 
Here, the turbulent heat flux $L_{T_{\mr{s}}}^{-1}Q_{\mr{s}}$ shown in Eqs. (9) and (11) is rewritten as $\eta_{\mr{s}}Q_{\mr{s}}$ 
by the use of the normalization with $L_{n_{0}}$. 
The deviation from the exact balance is also plotted by the dashed line, 
where one finds that the entropy balance relation is well satisfied for the whole simulation time in both the ITG and ETG cases. 
%
\begin{figure}
\centering
\includegraphics[scale=1.0]{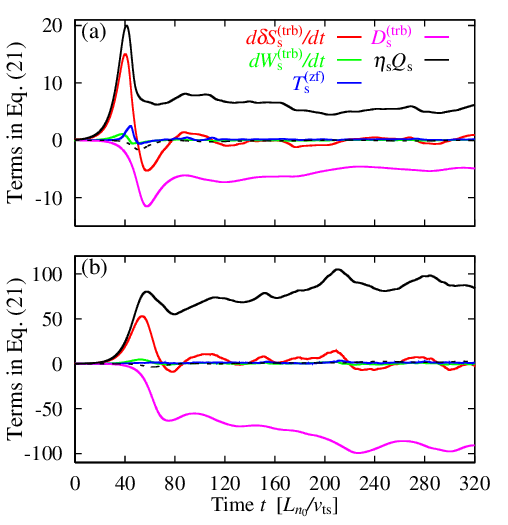}
\caption{(Color online) Time evolutions of each term in the entropy balance relation of the turbulence part, Eq. (21), 
for toroidal (a)ITG ($\mr{s}\! =\! \mr{i}$) and (b)ETG ($\mr{s}\! =\! \mr{e}$) turbulence. 
The deviation from the exact balance is also plotted by the dashed line. }
\end{figure}
%

As discussed in Sec. II, the turbulence part of the entropy $\delta S_{\! \mr{s}}^{\mr{(trb)}}$,
which characterizes the fluctuation intensity for the distribution function, is produced by the ITG (or ETG)-instability-driven 
heat flux $L_{T_{\mr{s}}}^{-1}Q_{\mr{s}}$.
Then, the saturation of the linear instability growth is observed at $t \! \sim \! 45$($\sim \! 60$) for the ITG (ETG) case. 
From the entropy balance equation for non-zonal modes [see Eq. (21)], one finds that the linear instability saturation 
is closely associated with both the nonlinear entropy transfer $\tzf$ and the collisional dissipation $D_{\mr{s}}^{\mr{(trb)}}$.
Thus, the quantity $\mathcal{R}_{\mr{s}} \! \equiv \! \tzf/(-D_{\mr{s}}^{\mr{(trb)}})$ provides ones with a good measure for
evaluating the effect of the nonlinear entropy transfer to zonal modes on the linear instability saturation. 
The typical values [evaluated at $t\! =\! 45$ (or $t \!= \!60$) for the ITG (or ETG) case] are
$\mathcal{R}_{\mr{i}} \! =\! 5.92 \times 10^{-1}$ and $\mathcal{R}_{\mr{e}} \! =\! 4.54 \times 10^{-2}$, respectively.
The larger value of $\mathcal{R}_{\mr{i}}$ indicates that the nonlinear entropy transfer from 
non-zonal ITG modes to zonal modes is significant for the saturation of the ITG instability growth, 
while the entropy transfer to zonal modes has little contribution to the saturation of the ETG instability growth. 
Indeed, some earlier works on the gyrokinetic simulations of toroidal ETG turbulence with adiabatic ions have also pointed out that 
the saturation of the ETG instability is mainly associated with the nonlinear coupling among non-zonal drift-wave fluctuations 
rather than the zonal-flow generation for the cases with moderate or strong magnetic-shear\cite{27,33}. 

As discussed in the previous section, $\mathcal{T}_{\! \mr{s}}^{\mr{(zf)}}$ is regarded as a kinetic extension of 
the zonal-flow production term due to the Reynolds stress, and works as a source for the zonal-flow component of the entropy 
$\delta {S}_{\mr{s}}^{\mr{(zf)}}$ and the potential energy $W_{\mr{s}}^{\mr{(zf)}}$ [see Eq. (22)]. 
In addition, it should be noted here that the zonal-flow generation is also affected by 
the coefficients of $\delta \phi_{k_{x},\,k_{y}=0}$ found in the right hand side of Eqs. (7) and (8) as well as those for 
$|\delta \phi_{k_{x},\,k_{y}=0}|^{2}$ in Eqs. (14) and (15) (One often refers to the coefficient as the ``zonal-flow inertia''), 
which are quite different between the ITG and ETG cases. 
In the simple evaluation with the long-wavelength limit of $\kp \rho_{\mr{ts}}^{2} \! \ll \! 1$, the zonal-flow inertia 
in the ITG and ETG turbulence are given by $\mathcal{M}_{\mr{i}} \! \equiv \! 1-\Gamma_{0\mr{i}} \! \simeq 
\! k_{x}^{2}\rho_{\mr{ti}}^{2}$ and $\mathcal{M}_{\mr{e}} \! \equiv \! 1+(T_{\mr{e}}/T_{\mr{i}})-\Gamma_{0\mr{e}} \! 
\simeq \! (T_{\mr{e}}/T_{\mr{i}})+k_{x}^{2}\rho_{\mr{te}}^{2}$, respectively. 
Then, for the typical wavenumber of $k_{x}\rho_{\mr{ts}} \! \sim \! 0.1$ observed in the turbulence simulations, 
the zonal-flow inertia for the ETG case is much larger than that for the ITG case: typically, 
$\mathcal{M}_{\mr{i}}/\mathcal{M}_{\mr{e}} \! \sim \! 10^{-2}$. 

In addition, one finds a time delay of the growth of collisional dissipation $|D_{\mr{i}}^{\mr{(trb)}}|$ in comparison with 
$\tzfi$ [cf. Fig. 2(a)]. 
This delay indicates that the entropy transfer to the higher-$k_{x}$ non-zonal modes mediated by the high-amplitude zonal flow, 
which is generated through the source $\tzfi$ and the small zonal-flow inertia $\mathcal{M}_{\mr{i}}$, 
occurs and enhances the subsequent collisional dissipation. 
This type of transfer processes will be discussed in detail in Sec. IV. 

It is also found that the statistically steady states are realized for $t \! \gtrsim \! 120$ in the ITG case 
and for $t \! \gtrsim \! 220$ in the ETG case, where the balance relation of 
$\eta_{\mr{s}}\overline{Q}_{\mr{s}} -\overline{\mathcal T}_{\! \mr{s}}^{\mr{(zf)}} \! \! = \! -\overline{D}_{\mr{s}}^{\mr{(trb)}}$ holds. 
(The time-average in the steady state is denoted by the overline.)  
The saturation levels of the turbulent heat flux for the ITG and ETG cases are evaluated as 
$\eta_{\mr{i}}\overline{Q}_{\mr{i}} \! = \! 5.31$ and $\eta_{\mr{e}}\overline{Q}_{\mr{e}} \! = \! 88.8$ in the unit of 
$(\rho_{\mr{ts}}/L_{n_{0}})^{2}(n_{0}\vv_{\mr{ts}})$, respectively, 
where the time average is taken over $ 220 \! \leqslant \! t \! \leqslant \! 320$ for both cases. 
As will be shown in Sec. III B, the strong zonal flows are sustained in the steady state of the ITG turbulence, 
while the radially elongated streamers with high amplitude are formed in the ETG case with the strong electron heat transport 
[cf. Figs. 4]. 

The similar plots for the entropy balance relation of the zonal-flow part, Eq. (22), are shown in Figs. 3(a) and 3(b). 
Also, the time-histories of the normalized entropy transfer $\tzf/\eta_{\mr{s}}\overline{Q}_{\mr{s}}$, 
which is useful for quantifying the effect of the entropy transfer on the transport level, are shown in Fig. 3(c). 
The deviation from the exact balance is quite small in both the ITG and ETG cases so that the linear and nonlinear dynamics 
of zonal flows are accurately solved.  
One also finds the balance relation of $\overline{\mathcal{T}}_{\! \mr{s}}^{\mr{(zf)}} \! \! = \! -\overline{D}_{\mr{s}}^{\mr{(zf)}}$ 
in the statistically steady states. 
%
\begin{figure}
\centering
\includegraphics[scale=1.0]{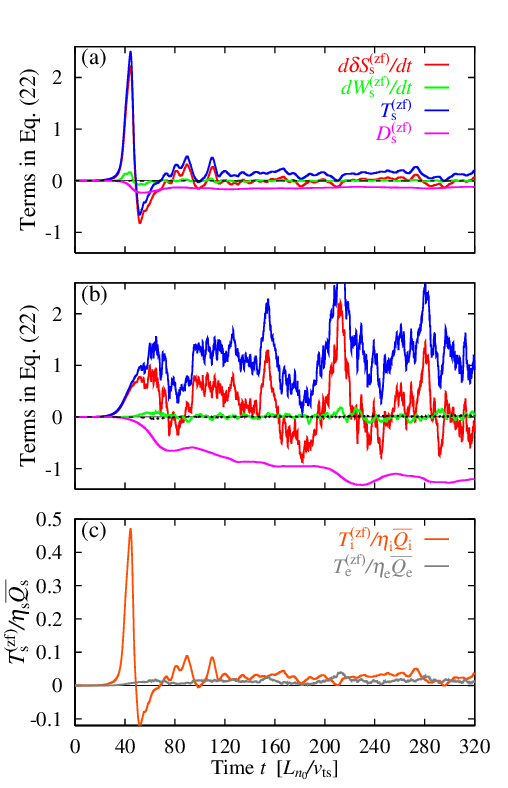}
\caption{(Color online) Time evolutions of each term in the entropy balance relation of the zonal-flow part, Eq. (22), 
for toroidal (a)ITG ($\mr{s}\! =\! \mr{i}$) and (b)ETG ($\mr{s}\! =\! \mr{e}$) turbulence. 
The deviation from the exact balance is also plotted by the dashed line. 
(c)The time-histories of the entropy transfer function normalized by the time-averaged heat flux $\tzf/\eta_{\mr{s}}\overline{Q}_{\mr{s}}$.} 
\end{figure}
%

A remarkable difference between the ITG and ETG cases is found in the time evolutions of $\tzf/\eta_{\mr{s}}\overline{Q}_{\mr{s}}$. 
The amplitude of $\tzfi/\eta_{\mr{i}}\overline{Q}_{\mr{i}}$ is much higher than 
$\tzfe/\eta_{\mr{e}}\overline{Q}_{\mr{e}}$ in the instability saturation phase of $t \! \sim \! 45$, 
where the entropy variable of non-zonal ITG modes is efficiently transferred to zonal modes. 
After the saturation of the linear instability, the amplitude of $\tzfi/\eta_{\mr{i}}\overline{Q}_{\mr{i}}$ decreases quickly 
to a low level which is of the same order of magnitude as (but still 1.8 times larger than) $\tzfe/\eta_{\mr{e}}\overline{Q}_{\mr{e}}$, 
where $\overline{\mathcal T}_{\! \mr{i}}^{\mr{(zf)}}/\eta_{\mr{i}}\overline{Q}_{\mr{i}} \!  = \! 2.56 \times 10^{-2}$ and 
$\overline{\mathcal T}_{\! \mr{e}}^{\mr{(zf)}}/\eta_{\mr{e}}\overline{Q}_{\mr{e}} \! = \! 1.42 \times 10^{-2}$ 
(the time-average is taken over $220 \! \leqslant \! t \! \leqslant \! 320$). 
The low amplitude of $\overline{\mathcal T}_{\! \mr{i}}^{\mr{(zf)}}/\eta_{\mr{i}}\overline{Q}_{\mr{i}}  \sim  
\overline{\mathcal T}_{\! \mr{e}}^{\mr{(zf)}}/\eta_{\mr{e}}\overline{Q}_{\mr{e}}$ indicates weak entropy transfer to zonal modes, 
which should balance with weak collisional dissipation for zonal modes. 

The comparison of time evolutions of $\tzfi/\eta_{\mr{i}}\overline{Q}_{\mr{i}}$ finds that, 
in contrast to the ETG case, significant entropy transfer from non-zonal to zonal modes contributes to the instability saturation, 
and then the small zonal-flow inertia give rise to the stronger zonal-flow generation in the ITG case. 
On the other hand, as seen from Figs. 2(a) and 2(b), the contribution of $\tzf$ is much smaller than $|D_{\mr{s}}^{\mr{(trb)}}|$ 
for both ITG and ETG cases (especially in the steady state) 
so that the difference of steady transport level is not explained seemingly in terms of $\tzf$. 
Indeed, a quite low amplitude of $\tzfi$ has also been observed in the steady state of ITG-TEM turbulence\cite{27}. 
These results suggest that the gross value of $\tzf$ (or Reynolds stress) does not characterize the role of zonal flows 
in the steady state so that one must further investigate the individual entropy transfer processes 
by means of the triad transfer function.
As will be shown in Sec. IV, different types of the entropy transfer processes are possible 
in the steady phases of the ITG and ETG turbulence. 

\subsection{Comparison of flow structures and heat-flux spectra}
A comparison of the vortex and flow structures, and the wavenumber spectra of turbulent heat flux in the steady phases of the ITG and ETG turbulence 
is given in this section. 
 
The vortex and flow structures in the three-dimensional flux-tube for the ITG and ETG turbulence are shown in Figs. 4(a) and 4(b), 
respectively, where the potential fluctuations $\delta \phi (x,y,z)$ at $t \! = \! 315$ are plotted.  
The strong zonal flows, which are translationally symmetric in the $y$- and $z$- directions, are observed in the ITG turbulence [Fig. 4(a)]. 
As shown in Sec. III A, the high-amplitude zonal flow is generated as a consequence of small zonal-flow inertia and efficient entropy transfer 
to zonal modes in the instability-saturation phase. 
On the other hand, radially elongated streamers with high amplitudes are sustained in the ETG turbulence [Fig. 4(b)]. 
One also finds that the streamers show ballooning-type structures which are similar to linear-mode structures, 
i.e., the amplitude on the perpendicular plane with $z \! =\! 0$ (the outboard side of the torus) decreases toward $z \! = \! \pm \pi$ 
(the inboard side of the torus). 
%
\begin{figure}
\centering
\includegraphics[scale=1.0]{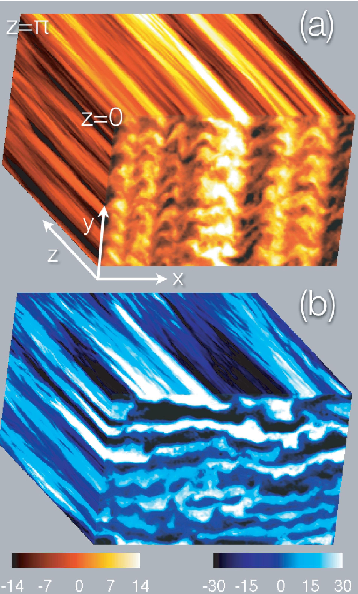}
\caption{(Color online) Contours of the potential fluctuations $\delta \phi(x,y,z)$ at $t \! = \! 315(L_{n_{0}}/\vv_{\mr{ts}})$ 
for toroidal (a)ITG and (b)ETG turbulence, where the unit is $(T_{\mr{s}}\rho_{\mr{ts}}/eL_{n_{0}})$. 
The box size is $L_{x}  \times  L_{y}  \times L_{z} \! = \! 266 \rho_{\mr{ts}}  \times  168\rho_{\mr{ts}}  \times \pi$. 
The ($x,\,y$)-cross section shown here is the perpendicular plane in the outboard side of the torus, where $z \! = \! 0$.}
\end{figure}
%

Transport properties in the steady states of the ITG and ETG turbulence are compared in terms of the wavenumber spectra of 
the turbulent heat flux $\eta_{\mr{s}}\overline{Q}_{\mr{s}\subp{k}}$ in the two-dimensional $\bm{k}_{\perp}$-space which are plotted 
in Figs. 5(a) and 5(b).
For reference and comparison, the wavenumber spectra of potential fluctuations are also plotted in Figs. 5(c) and 5(d), 
where each amplitude is normalized with the maximum value in the non-zonal components ($k_{y} \! \neq \! 0$). 
While the mode with ($k_{x}\! \simeq \! 0$, $k_{y} \! \simeq \! 0.2\rho_{\mr{ts}}^{-1}$) makes the largest contribution to the heat transport 
in both cases, one clearly finds a qualitative difference in the spectra of the turbulent heat flux and potential fluctuations between 
the ITG and ETG turbulence.
The wavenumber spectra for the ITG turbulence significantly expands into the high $k_{x}$-region [Figs. 5(a) and 5(c)] where 
the modes are stabilized [cf. Fig. 1] and make less contribution to the heat transport, 
while the spectra for the ETG case is confined in the lower-$k_{x}$ region. 
The elongated and confined structures of the wavenumber spectra observed in the ITG and ETG cases reflect their different entropy transfer processes 
in the steady phases, as will be discussed in detail in the next section. 
%
\begin{figure*}
\centering
\includegraphics[scale=1.0]{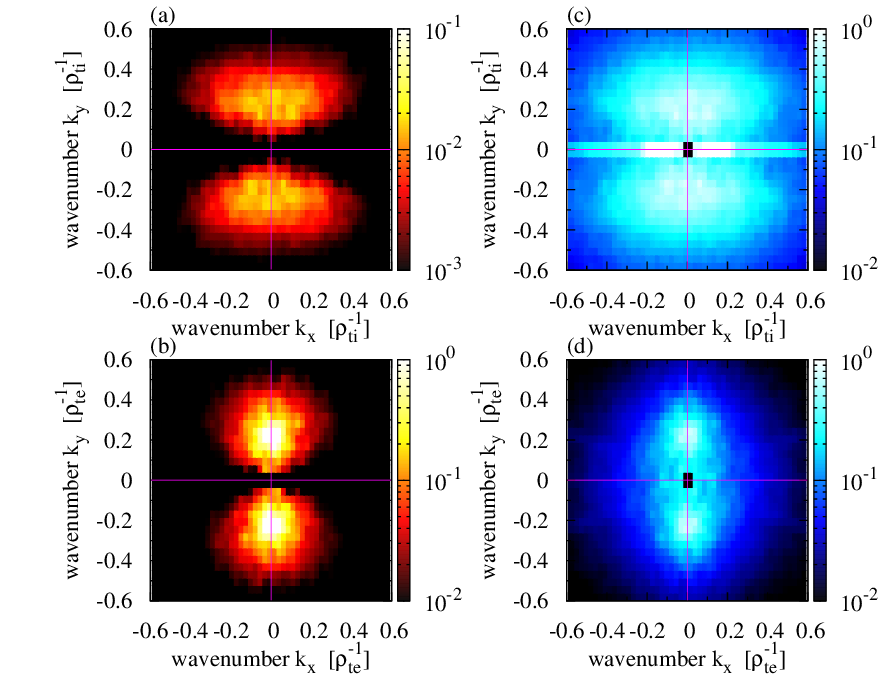}
\caption{(Color online) Wavenumber spectra of the turbulent heat flux $\eta_{\mr{s}}Q_{\mr{s}\subp{k}}$ [(a) and (b)] and
the potential fluctuation $\langle |\dphik| \rangle $ [(c) and (d)] 
in the steady states of the ITG (upper row) and ETG (lower row) turbulence, 
where the amplitudes are averaged over $220 \! \leqslant \! t \! \leqslant 320$. 
The amplitudes of potential fluctuations are normalized by the maximum value in the non-zonal components. }
\end{figure*}
%

\section{NONLINEAR ENTROPY TRANSFER VIA ZONAL MODES} 
In this section, nonlinear interactions between zonal flows and the ambient turbulence in the instability-saturation and steady phases 
are addressed based on the spectral analysis of the triad entropy transfer function given in Sec. II. 

In the previous section, the entropy balance relations for non-zonal and zonal modes in the ITG and ETG turbulence have been discussed 
(cf. Figs. 2 and 3). 
Particularly, the different time-evolution of the normalized entropy transfer function $\tzf/\eta_{\mr{s}}\overline{Q}_{\mr{s}}$ 
has been shown [cf. Fig. 3(c)]. 
The amplitude of $\tzfi/\eta_{\mr{i}}\overline{Q}_{\mr{i}}$ is much higher than 
$\tzfe/\eta_{\mr{e}}\overline{Q}_{\mr{e}}$ in the saturation phase of the linear instability. 
In the steady state, we observe the lower amplitude of $\tzfi/\eta_{\mr{i}}\overline{Q}_{\mr{i}}$ with 
the same order of magnitude as $\tzfe/\eta_{\mr{e}}\overline{Q}_{\mr{e}}$, which should balance with weak collisional 
dissipation for zonal modes. 
It should be stressed here that, even though there are little entropy transfer to the zonal modes in the steady state, 
the ITG-driven high-amplitude zonal flows still play an important role in the transport suppression through a different type of 
the entropy transfer process from that in the saturation phase, as shown below. 
In order to understand more accurately the nonlinear entropy transfer processes in the saturation and steady states,
the spectral analysis of the triad entropy transfer function $\mathcal{J}_{\mr{s}}[ \bm{k}_{\perp}| \bm{p}_{\perp}, \bm{q}_{\perp} ]$ 
(rather than $\tzf$) with the aid of the detailed balance relation shown in Eq. (20) is necessary, 
because $\tzf$ (or, equivalently, $\mathcal{T}_{\mr{s}\subp{k}}$) obscures individual entropy transfer processes 
among the non-zonal modes via the zonal mode, and provides only the net amount of the entropy transferred from 
non-zonal (turbulence) to zonal modes. 
The detailed spectral analysis of the triad transfer function can reveal the critical role of zonal flows 
in the nonlinear entropy transfer processes, which are closely associated with the saturation of instability growth 
and transport suppression in the steady state. 

Hereafter, we consider the entropy transfer processes associated with the nonlinear interactions among two non-zonal modes 
with $\bm{p}_{\perp}$ and $\bm{q}_{\perp}$ and a zonal mode with $\bm{k}_{\mr{zf}} \! = \! k_{\mr{zf}} \nabla x$, 
which satisfy the triad-interaction condition $\bm{k}_{\mr{zf}} \! + \bm{p}_{\perp} \! + \bm{q}_{\perp} \! = 0$. 
First, the non-zonal primary mode with $\bm{p}_{\perp}$ is chosen to be the ``transport-driving mode'' 
with ($k_{x}\! \simeq \! 0$, $k_{y} \! \simeq \! 0.2\rho_{\mr{ts}}^{-1}$) which makes the most dominant contribution to the turbulent 
heat flux, as shown in Figs. 5(a) and 5(b). 

The wavenumber spectrum of the triad entropy transfer function normalized by the time-averaged heat flux, i.e., 
$\overline{\mathcal{J}}_{\mr{s}}[ \bm{k}_{\mr{zf}}| \,\bm{p}_{\perp}, \bm{q}_{\perp} ]/\eta_{\mr{s}}\overline{Q}_{\mr{s}}$, 
in the saturation phases of ITG and ETG instabilities are shown in Figs. 6(a) and 6(b), respectively, 
where the time-average is taken over $30 \! \leqslant \! t \! \leqslant \! 45$. 
Here, the wavenumbers of the ITG- and ETG-driven zonal flows are, respectively, set to $k_{\mr{zf}} \! =\! 0.1410 \rho_{\mr{ti}}^{-1}$ 
and $k_{\mr{zf}} \! = \! 0.0705 \rho_{\mr{te}}^{-1}$, which have the largest amplitude in the zonal-flow components. 
The spectrum is symmetric with respect to the point $(q_{x},\, q_{y}) \! =\! (-k_{\mr{zf}},\, 0)$ 
due to the symmetric property of the triad transfer function for the zonal mode with $k_{\mr{zf}}$. 
In the ITG turbulence [Fig. 6(a)], one clearly finds that the large positive values of 
$\overline{\mathcal{J}}_{\mr{i}}[ \bm{k}_{\mr{zf}}| \,\bm{p}_{\perp}, \bm{q}_{\perp} ]/\eta_{\mr{i}}\overline{Q}_{\mr{i}}$ 
spread over the linearly unstable region around $q_{y} \! \simeq \! 0.4\rho_{\mr{ti}}^{-1}$ [cf. Fig. 1]. 
The entropy transfer from the linearly unstable non-zonal modes to zonal modes significantly contributes 
to the saturation of the ITG instability. 
On the other hand, the lower amplitude of 
$\overline{\mathcal{J}}_{\mr{e}}[ \bm{k}_{\mr{zf}}| \,\bm{p}_{\perp}, \bm{q}_{\perp} ]/\eta_{\mr{e}}\overline{Q}_{\mr{e}}$ 
is observed in the ETG case [Fig. 6(b)], 
where $(\overline{\mathcal{J}}_{\mr{e}}/\eta_{\mr{e}}\overline{Q}_{\mr{e}})_{\mr{max}}/ 
(\overline{\mathcal{J}}_{\mr{i}}/\eta_{\mr{i}}\overline{Q}_{\mr{i}})_{\mr{max}} \simeq 0.431$. 
It is thus found that the generation of ETG-driven zonal flows in the saturation process is less effective than that in the ITG case. 
These results are consistent with the different amplitude of $\tzfi/\eta_{\mr{i}}\overline{Q}_{\mr{i}}$ and 
$\tzfe/\eta_{\mr{e}}\overline{Q}_{\mr{e}}$ observed in the saturation phase, as shown in Fig. 3(c). 
%
\begin{figure}
\centering
\includegraphics[scale=1.0]{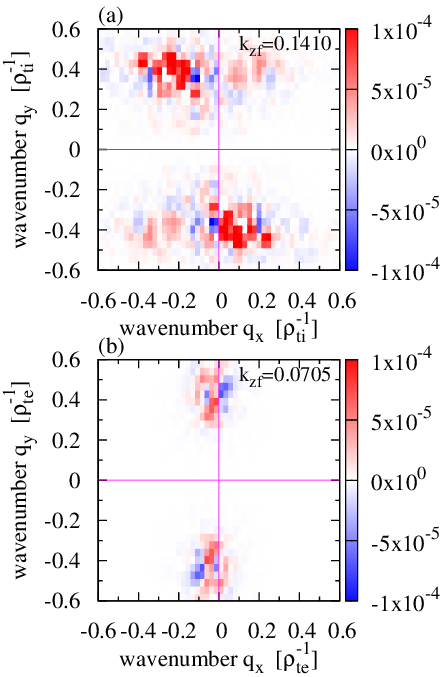}
\caption{(Color online) Wavenumber spectrum of the triad transfer function normalized by the mean heat flux, 
$\overline{\mathcal{J}}_{\mr{s}}[ \bm{k}_{\mr{zf}}| \,\bm{p}_{\perp}, \bm{q}_{\perp} ]/\eta_{\mr{s}}\overline{Q}_{\mr{s}}$, 
for the fixed-$\bm{k}_{\mr{zf}}$ in the saturation phase of toroidal (a)ITG ($\mr{s}\! =\! \mr{i}$) 
and (b)ETG ($\mr{s}\! =\! \mr{e}$) turbulence, where the time-average is taken over $30 \! \leqslant \! t \! \leqslant \! 45$.}
\end{figure}
%

In the steady state, the entropy transfer processes are qualitatively different from those in the saturation phase. 
The wavenumber spectra of 
$\overline{\mathcal{J}}_{\mr{s}}[ \bm{k}_{\mr{zf}}| \,\bm{p}_{\perp}, \bm{q}_{\perp} ]/\eta_{\mr{s}}\overline{Q}_{\mr{s}}$ 
in the steady state are shown in Figs. 7. 
It is observed in Fig. 7(a) that the amplitude of 
$\overline{\mathcal{J}}_{\mr{i}}[ \bm{k}_{\mr{zf}}| \,\bm{p}_{\perp}, \bm{q}_{\perp} ]/\eta_{\mr{i}}\overline{Q}_{\mr{i}}$ 
is quite low in comparison to that in the saturation phase shown above. 
For the ETG case [Fig. 7(b)], the large positive values of 
$\overline{\mathcal{J}}_{\mr{e}}[ \bm{k}_{\mr{zf}}| \,\bm{p}_{\perp}, \bm{q}_{\perp} ]/\eta_{\mr{e}}\overline{Q}_{\mr{e}}$ 
are found near the region of transport-driving modes [cf. Fig. 5(b)], but are partly canceled by the large negative values. 
Note that, as discussed above, the generation of ETG-driven zonal flows is less effective due to the large zonal-flow 
inertia, even if the relatively large positive values of 
$\overline{\mathcal{J}}_{\mr{e}}[ \bm{k}_{\mr{zf}}| \,\bm{p}_{\perp}, \bm{q}_{\perp} ]/\eta_{\mr{e}}\overline{Q}_{\mr{e}}$ 
are observed in comparison to those in the ITG case. 
%
\begin{figure}
\centering
\includegraphics[scale=1.0]{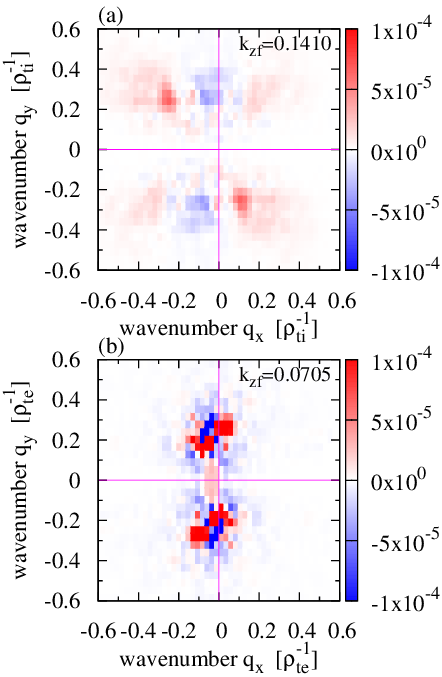}
\caption{(Color online) Wavenumber spectrum of the triad transfer function normalized by the mean heat flux, 
$\overline{\mathcal{J}}_{\mr{s}}[ \bm{k}_{\mr{zf}}| \,\bm{p}_{\perp}, \bm{q}_{\perp} ]/\eta_{\mr{s}}\overline{Q}_{\mr{s}}$, 
for the fixed-$\bm{k}_{\mr{zf}}$ in the steady state of toroidal (a)ITG ($\mr{s}\! =\! \mr{i}$) 
and (b)ETG ($\mr{s}\! =\! \mr{e}$) turbulence, where the time-average is taken over $220 \! \leqslant \! t \! \leqslant \! 320$.}
\end{figure}
%

Although the entropy transfer to zonal modes is quite weak in the steady state of the ITG turbulence, 
i.e., $\overline{\mathcal{J}}_{\mr{i}}[ \bm{k}_{\mr{zf}}| \,\bm{p}_{\perp}, \bm{q}_{\perp} ] \! \simeq \! 0$, 
the ITG-driven zonal flows still work as a `` mediator'' on the entropy transfer from non-zonal transport-driving modes 
to other non-zonal modes with higher radial-wavenumbers which make less contribution to the turbulent heat flux [cf. Fig. 1 and 5(a)]. 
In order to clarify this point, the wavenumber spectra of 
$\overline{\mathcal{J}}_{\mr{i}}[ \bm{p}_{\perp}| \,\bm{q}_{\perp}, \bm{k}_{\mr{zf}} ]/\eta_{\mr{i}}\overline{Q}_{\mr{i}}$ 
in the steady state for three different $\bm{p}_{\perp}$'s is plotted in Figs. 8(a) -- (c), 
where $p_{y}$ is fixed to $p_{y} \! = \! 0.2250\rho_{\mr{ti}}^{-1}$ which is the dominant component driving the heat flux. 
It is clearly found that all the figures commonly show the two-stripe pattern at $p_{y} \! = \! 0$ and $p_{y} \! =\! -0.2250\rho_{\mr{ti}}^{-1}$ 
which indicates that the $\bm{p}_{\perp}$-mode dominantly interacts with other non-zonal modes through the zonal modes. 
Note that, in general, one can not uniquely determine the direction of the entropy transfer between ``two modes'' 
only from the sign of $\overline{\mathcal{J}}_{\mr{i}}[ \bm{p}_{\perp}| \,\bm{q}_{\perp}, \bm{k}_{\mr{zf}} ]$ 
because the ``triad'' transfer function describes the entropy transfer among ``three modes forming a triad interaction''. 
However, the direction of the entropy transfer between two non-zonal modes can be identified in the present case as 
$\bm{p}_{\perp} \! \rightarrow \bm{q}_{\perp}$ for negative $\overline{\mathcal{J}}_{\mr{i}}[ \bm{p}_{\perp}| \,\bm{q}_{\perp}, \bm{k}_{\mr{zf}} ]$, 
or as $\bm{q}_{\perp} \! \rightarrow \bm{p}_{\perp}$ 
for positive one due to the detailed balance relation given in Eq. (20) leading to 
%
\begin{equation}
\overline{\mathcal{J}}_{\mr{i}}\left [ \bm{p}_{\perp}| \bm{q}_{\perp}, \bm{k}_{\mr{zf}} \right ] 
+\overline{\mathcal{J}}_{\mr{i}}\left [ \bm{q}_{\perp}| \bm{k}_{\mr{zf}}, \bm{p}_{\perp} \right ] 
=-\overline{\mathcal{J}}_{\mr{i}}\left [ \bm{k}_{\mr{zf}}| \bm{p}_{\perp}, \bm{q}_{\perp} \right ] \simeq 0\ , 
\end {equation}
%
where the last equality is confirmed by the small amplitude of 
$\overline{\mathcal{J}}_{\mr{i}}[ \bm{k}_{\mr{zf}}| \,\bm{p}_{\perp}, \bm{q}_{\perp} ]$ in the steady state as shown in Fig. 7(a). 

The entropy transfer processes shown in Figs. 8(a) -- (c) are explained more in detail as follows:
From Fig. 8(a), we observe the large negative values at the zonal mode with 
$q_{x}\! =\! k_{\mr{zf}} \! = \! 0.1410 \rho_{\mr{ti}}^{-1}$ and $q_{y} \! = \! 0$ [represented by solid green(light gray) arrow] and 
at the non-zonal mode with 
$q_{x}\! =\! -k_{zf}$ and $q_{y} \! = \! -p_{y} \! =\! -0.2250 \rho_{\mr{ti}}^{-1}$ [solid blue(darker gray) arrow], 
which form a triad with the transport-driving primary mode with $\bm{p}_{\perp}$ [solid red(gray) arrow]. 
This means that the entropy is transferred from the non-zonal primary mode ($\bm{p}_{\perp}$) to 
the higher radial-wavenumber mode ($\bm{q}_{\perp}$) through the triad interaction with high-amplitude zonal mode ($\bm{k}_{\mr{zf}}$). 
Besides, the $-\bm{q}_{\perp}$-mode, which is equivalent to the $\bm{q}_{\perp}$-mode due to the reality of the physical variable, 
is plotted by the dotted blue(darker gray) arrow in the figure. 
In Fig. 8(b), the previously transferred mode [dotted blue(darker gray) arrow in Fig. 8(a)], which consists of a triad with two modes shown by solid black 
arrows [equivalent to solid red(gray) and green(light gray) arrows in Fig. 8(a)], is now considered as the new primary mode with 
$p_{x} \! = \! 0.1410\rho_{\mr{ti}}^{-1}$ and $p_{y} \! =\! 0.2250\rho_{\mr{ti}}^{-1}$ shown by solid red(gray) arrow. 
Then, the entropy variable is further transferred to the non-zonal mode with higher radial-wavenumber 
(solid, or equivalently, dotted blue(darker gray) arrow) via the interaction with the zonal mode with 
$k_{\mr{zf}} \! \simeq \! 0.1410\rho_{\mr{ti}}^{-1}$ (shown by solid green(light gray) arrow).  
The ``successive'' entropy transfer to the higher radial-wavenumber mode is also found in Fig. 8(c). 
[One again finds that the entropy of the primary mode (solid red(gray) arrow) is coming through the interaction with two modes 
(solid black arrows), and is going to the higher radial-wavenumber mode (solid blue(darker gray) arrows) via the interaction with the zonal mode 
(solid green(light gray) arrow).]
%
\begin{figure}
\centering
\includegraphics[scale=1.0]{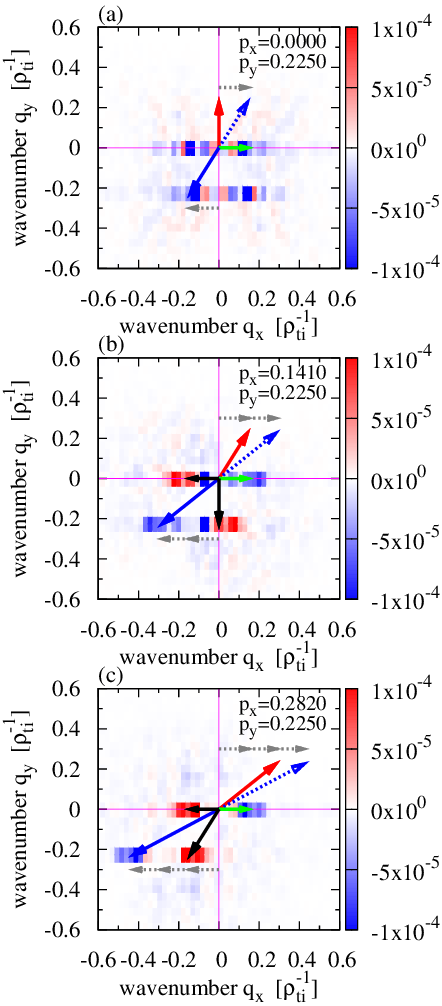}
\caption{(Color online) Wavenumber spectra of the triad transfer function normalized by the mean heat flux, 
$\overline{\mathcal{J}}_{\mr{i}}[ \bm{p}_{\perp}| \,\bm{q}_{\perp}, \bm{k}_{\mr{zf}} ]/\eta_{\mr{i}}\overline{Q}_{\mr{i}}$, 
for three different $\bm{p}_{\perp}$'s with $p_{y} \! = \! 0.2250 \rho_{\mr{ti}}^{-1}$(fixed) in the steady state of toroidal 
ITG turbulence, where the time-average is taken over $220 \! \leqslant \! t \! \leqslant \! 320$. }
\end{figure}
%

The spectral analysis of the triad transfer function reveals that, through the triad interaction with the high-amplitude zonal flows 
in the wavenumber-space, the entropy of the non-zonal transport-driving modes is successively transferred to the non-zonal modes 
with higher radial-wavenumbers for which stronger damping occurs due to the finite gyroradius effect. 
These results provide ones with a novel physical picture of the suppression of the ITG turbulent transport by zonal flows 
in the steady state, from the view point of the gyrokinetic entropy balance and transfer. 

Figures 9(a) -- (c) show the wavenumber spectra of 
$\overline{\mathcal{J}}_{\mr{e}}[ \bm{p}_{\perp}| \,\bm{q}_{\perp}, \bm{k}_{\mr{zf}} ]/\eta_{\mr{e}}\overline{Q}_{\mr{e}}$ 
in the steady state of toroidal ETG turbulence for three different $\bm{p}_{\perp}$'s, respectively. 
In contrast to the ITG turbulence, no remarkable spectral structures suggesting the successive entropy transfer to 
the non-zonal modes with higher radial-wavenumbers are observed. 
Instead, the entropy transfer within a low wavenumber region occurs dominantly 
through the nonlinear interactions among non-zonal modes. 
One also finds the net entropy transfer to the primary mode with $\bm{p}_{\perp}$ through the coupling 
of zonal flows and radially elongated streamers with $q_{x} \! \simeq \! 0$ [cf. Fig. 9(c)], 
while the subsequent transfer to non-zonal mode with the higher-radial wavenumber does not occur. 
Thus, in both the saturation and steady phases of the ETG turbulence, 
the role of zonal flows in the successive entropy transfer to the higher wavenumber modes is much weaker than that in the ITG case. 
%
\begin{figure}
\centering
\includegraphics[scale=1.0]{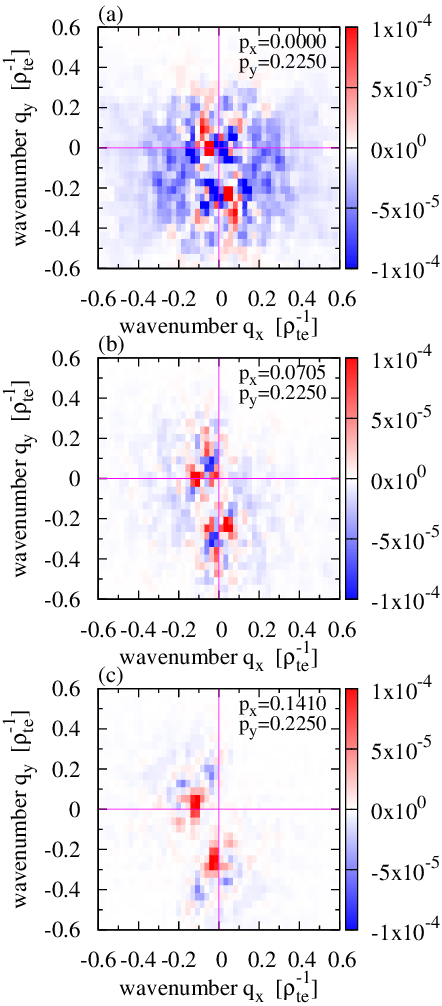}
\caption{(Color online) Wavenumber spectra of 
$\overline{\mathcal{J}}_{\mr{e}}[ \bm{p}_{\perp}| \,\bm{q}_{\perp}, \bm{k}_{\mr{zf}} ]/\eta_{\mr{e}}\overline{Q}_{\mr{e}}$, 
for three different $\bm{p}_{\perp}$'s with $p_{y} \! = \! 0.2250 \rho_{\mr{te}}^{-1}$(fixed) in the steady state of toroidal 
ETG turbulence, where the time-average is taken over $220 \! \leqslant \! t \! \leqslant \! 320$. }
\end{figure}
%

In addition to the individual entropy transfer processes involving the interactions with all components of zonal modes shown above, 
the wavenumber spectra of 
$\overline{\mathcal{J}}_{\mr{s}}[ \bm{p}_{\perp}| \,\bm{q}_{\perp}, \bm{k}_{\mr{zf}} ]/\eta_{\mr{s}}\overline{Q}_{\mr{s}}$ are plotted in Figs. 10 
as a function of $p_{x}$ and $q_{x}$, where $p_{y} \! =\! -q_{y} \! = \! 0.2250\rho_{\mr{ts}}^{-1}$ and $k_{\mr{zf}}$ giving the largest 
amplitude in the zonal-flow components are fixed. 
In the ITG case [Fig. 10(a)], one clearly finds the elongated diagonal stripes indicating that the entropy at $p_{x} \! =\! 0$ 
(transport-driving mode) is transferred to the higher radial-wavenumber non-zonal modes via the successive interactions 
with zonal modes. 
The successive transfer processes are represented by the dotted gray arrows in Fig. 10(a). 
In contrast, in the ETG case [Fig. 10(b)], the elongated diagonal structures indicating the successive transfer are no longer observed.   
%
\begin{figure}
\centering
\includegraphics[scale=1.0]{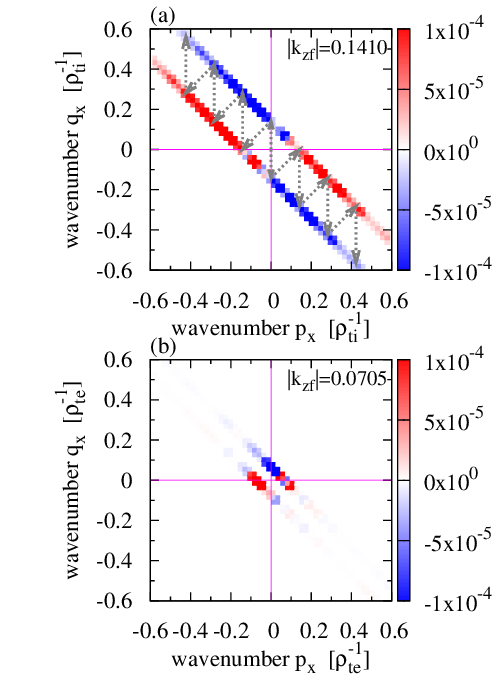}
\caption{(Color online) Wavenumber spectrum of 
$\overline{\mathcal{J}}_{\mr{s}}[ \bm{p}_{\perp}| \,\bm{q}_{\perp}, \bm{k}_{\mr{zf}} ]/\eta_{\mr{s}}\overline{Q}_{\mr{s}}$ for fixed-$|k_{\mr{zf}}|$ 
with $p_{y} \! =\! -q_{y} \! = \! 0.2250\rho_{\mr{ts}}^{-1}$ in the steady phases of toroidal (a)ITG ($\mr{s}\! =\! \mr{i}$) and 
(b)ETG ($\mr{s}\! =\! \mr{e}$) turbulence, where the time-average is taken over $220 \! \leqslant \! t \! \leqslant \! 320$. 
The lower and upper lines in each figure correspond to $k_{\mr{zf}}+p_{x}+q_{x}=0$ for $k_{\mr{zf}}\! > \! 0$ and $k_{\mr{zf}} \! < \! 0$, 
respectively.}
\end{figure}
%

Finally, in Figs. 11, we briefly summarize the nonlinear entropy transfer processes in the saturation and the steady phases of 
the toroidal ITG and ETG turbulence. 
In the saturation phase of the toroidal ITG instability growth, 
the entropy of non-zonal ITG modes is significantly transferred to zonal modes[Fig. 11(a)]. 
In the steady state of the ITG turbulence, the entropy transfer to zonal modes becomes quite weak, i.e., 
$\tzfi/\eta_{\mr{i}}\overline{Q}_{\mr{i}} \! \sim \! \tzfe/\eta_{\mr{e}}\overline{Q}_{\mr{e}} \! \ll \! 1$. 
Nevertheless, the ITG-driven zonal flows mediate the entropy transfer process from non-zonal modes 
strongly driving the heat transport to the higher radial-wavenumber (non-zonal) modes with less contribution to the turbulent 
heat flux [Fig. 11(b)]. 
The successive entropy transfer to the higher radial-wavenumber modes due to the triad interaction with the strong zonal flows 
is associated with the broadening of the spectra of potential fluctuations and heat flux 
in the $k_{x}$-direction shown in Figs. 5(a) and 5(c). 
The successive entropy transfer via the high-amplitude zonal flow also occurs in the saturation phase, 
and enhances the subsequent collisional dissipation. 
%
\begin{figure*}
\centering
\includegraphics[scale=1.0]{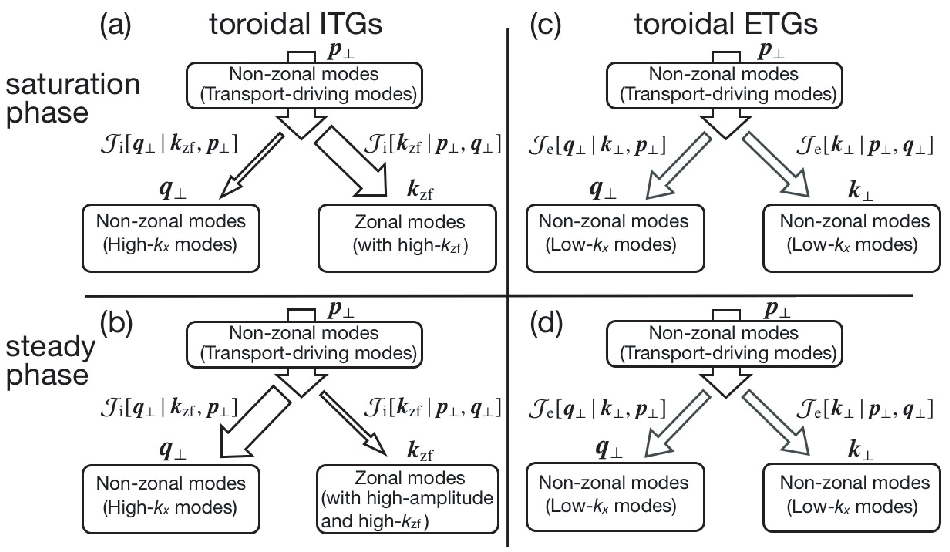}
\caption{Summary of the entropy transfer processes in the saturation (upper row) and the steady (lower row) phases 
for the toroidal ITG (left column) and ETG (right column) turbulence, where arrows denote the direction of the entropy transfer.}
\end{figure*}
%

In contrast to the ITG case, the nonlinear interactions among the low-wavenumber non-zonal modes including the radially elongated 
streamers are dominant in the ETG case so that the entropy transfer resulting from the interactions with zonal flows are not 
effective for both the saturation of the instability growth and the transport regulation in the steady state [Figs. 11(c) and 11(d)].
As pointed out in some earlier works on the toroidal ETG turbulence\cite{10}, the vortex structures and zonal flows strongly 
depend on the geometrical and plasma parameters. 
In particular, strong ETG-driven zonal flows are observed in the case with weak (or negative) magnetic shear, 
where the electron heat transport is significantly reduced. 
It is not trivial whether the entropy transfer function for the toroidal ETG turbulence with the strong zonal flows 
shows the similar wavenumber spectrum to that in the ITG case presented here. 
The entropy transfer analysis for the toroidal ETG turbulence with weak magnetic-shear remains as a future work. 

Comparisons of the entropy transfer processes between the slab and toroidal systems have also been discussed in detail in Ref. 34, 
and have revealed that the significant entropy transfer from non-zonal to zonal modes is also observed in the saturation phases of 
slab ITG and ETG instabilities. 
(Note that the zonal-flow generation in the slab ETG case is stronger than that in the toroidal one 
because there is no neoclassical shielding effects for the zonal-flow potential.)
Similar to the toroidal ITG case, the entropy transfer to zonal modes are quite weak in the steady state in the slab ITG turbulence,
and then the successive transfer to the higher radial-wavenumber modes is also found.
In the steady state of the slab ETG turbulence with turbulent vortices driving the steady heat transport [cf. Ref. 11], 
it is found that the entropy transfer dominantly occurs in the low wavenumber region through the nonlinear interactions 
among non-zonal modes, which is similar to that of the toroidal ETG case. 

\section{CONCLUDING REMARKS}
In the present study, the entropy balance relations for non-zonal and zonal modes in toroidal ITG and ETG turbulence 
have been examined by means of five-dimensional nonlinear gyrokinetic Vlasov simulation code, GKV\cite{2}. 
Particularly, the nonlinear entropy transfer processes in the saturation and steady phases, which are associated with the 
instability saturation and the transport regulation, are investigated by the spectral analysis of the triad entropy transfer function 
with the aid of the detailed balance relation. 

The nonlinear simulation results confirm that the entropy balance relations for turbulence (non-zonal) and zonal-flow components, 
which are coupled with each other through the entropy transfer function $\tzf$, are accurately satisfied for the whole simulation 
time both in the ITG and ETG turbulence. 
It is also found that the statistically steady states of the ITG and ETG turbulence are realized long after the saturation of linear 
instability, where the balance relations of 
$\overline{\mathcal T}_{\! \mr{s}}^{\mr{(zf)}} \! \!  = \! -\overline{D}_{\mr{s}}^{\mr{(zf)}} \! \geqslant \! 0$ 
and $L_{T_{\mr{s}}}^{-1}\overline{Q}_{\mr{s}}-\overline{\mathcal T}_{\! \mr{s}}^{\mr{(zf)}}\! \! =\! -\overline{D}_{\mr{s}}^{\mr{(trb)}}$ 
hold separately for the turbulence and zonal-flow components (the overline denotes the time-average in the steady state). 
A remarkable difference between the ITG and ETG cases is found in the time evolutions of $\tzf/\eta_{\mr{s}}\overline{Q}_{\mr{s}}$. 
In the saturation phase of the instability growth, the higher peak amplitude of $\tzfi/\eta_{\mr{i}}\overline{Q}_{\mr{i}}$ 
in comparison to that of $\tzfe/\eta_{\mr{e}}\overline{Q}_{\mr{e}}$ is observed, 
while it decreases significantly in the steady state and becomes same order of magnitude as that for the ETG case, i.e.,  
$\tzfi/\eta_{\mr{i}}\overline{Q}_{\mr{i}} \! \sim \! \tzfe/\eta_{\mr{e}}\overline{Q}_{\mr{e}} \! \ll \! 1$. 
The significant entropy transfer to zonal modes, which leads to the high-amplitude zonal flows, 
play a critical role in the saturation of the ITG instability growth whereas the entropy transfer processes among non-zonal modes are dominant in the ETG case. 

In the steady state, the strong zonal flows are sustained in ITG case, while the radially elongated streamers with the high amplitudes, 
which are associated with the strong electron heat transport, are formed in the ETG case.
Although the mode with $k_{x}\! \simeq \! 0$ and $k_{y} \! \simeq \! 0.2\rho_{\mr{ts}}^{-1}$ makes the most dominant contribution to 
the heat transport, different spectral shapes of the heat flux and the potential fluctuations are found between the ITG and ETG turbulence, that is, 
broad $k_{x}$-spectra are observed for the ITG case while the spectra for the ETG case are confined within the lower-$k_{x}$ region. 

In order to understand the detailed entropy transfer processes in the saturation and steady phases,
the wavenumber spectra of the triad transfer functions $\mathcal{J}_{\mr{s}}[ \bm{k}_{\mr{zf}}|\,\bm{p}_{\perp}, \bm{q}_{\perp} ]$ and 
$\mathcal{J}_{\mr{s}} [ \bm{p}_{\perp}| \bm{q}_{\perp}, \bm{k}_{\mr{zf}} ]$ 
for two non-zonal modes with $\bm{p}_{\perp}$ and $\bm{q}_{\perp}$ and a zonal mode with 
$\bm{k}_{\mr{zf}} \! = \! k_{\mr{zf}} \nabla x$ have been examined. 
The detailed balance relation, Eq. (20) [or Eq. (23)], provides us explicitly with the direction of the entropy transfer among 
the zonal and non-zonal modes. 
Then, the qualitatively different entropy transfer processes in the saturation and steady phases of the ITG turbulence are 
clarified. 
The entropy transfer from non-zonal to zonal modes is substantial in the saturation phase of the ITG turbulence, 
while, once the high-amplitude zonal flows are generated, the entropy transfer to the zonal modes becomes quite weak 
in the steady state. 
Instead, the zonal flows mediate the entropy transfer among non-zonal modes, i.e., 
the entropy of non-zonal modes with low radial-wavenumbers driving the heat transport is successively transferred 
to the other non-zonal modes with higher radial-wavenumbers with less contribution to the transport. 
The successive entropy transfer to the high-$k_{x}$ modes due to the triad interactions with the strong zonal flows 
in the wavenumber-space leads to the broadening of the wavenumber spectra in the steady state of the ITG turbulence. 
In contrast to the ITG case, the nonlinear interactions among the low-wavenumber non-zonal modes are dominant in the ETG case 
so that the entropy transfer resulting from the interactions with zonal flows are not 
effective for the saturation of the instability growth and the transport regulation. 

The results obtained from the triad entropy transfer analysis introduced here 
provide ones with a new understanding of the nonlinear interaction between turbulence and 
zonal flows and their direct effects on the turbulent transport level. 
These methods can be applied to general plasma turbulent transport problems, e.g., ITG-TEM driven turbulent transport. 
Moreover, the entropy transfer analysis may give us useful suggestions for advanced turbulence diagnostics such as the bi-spectrum analysis for experimental data, 
because it involves the explicit information of the direction of the transfer which has rarely been discussed in the conventional analysis. 

\begin{acknowledgments}
Numerical computations are performed on the NIFS Plasma Simulator. 
This work is supported in part by the Grant-in-Aid for Japan Society for the Promotion of Science Fellowship (No.20-4017), 
and in part by the Japanese Ministry of Education, Culture, Sports, Science, and Technology, Grant No. 21560861, 
and in part by the NIFS Collaborative Research Programs, NIFS10KTAT040, NIFS10KNTT003, and NIFS10KNXN186. 
\end{acknowledgments}
%

\appendix*
\section{SYMMETRIZED TRIAD TRANSFER FUNCTION}
A formal derivation of the symmetrized triad transfer function described in Eq. (17) is shown in this Appendix. 

Let us describe the gyrokinetic equation [Eq. (1)] symbolically as a general dynamical equation with a quadratic nonlinearity, 
%
\begin{equation}
\lbr \pdif{}{t} - \mathcal{L}_{\sub{k}} \rbr X_{\sub{k}} + \sum_{\bm{p}} \sum_{\bm{q}} \mathcal{M}_{\bm{k}}^{\bm{p},\bm{q}} 
Y_{\sub{p}}^{\ast}X_{\sub{q}}^{\ast}\delta_{\bm{k}+\bm{p}+\bm{q},0} = 0\ , 
\end{equation}
%
where $X_{\sub{k}}$ and $Y_{\sub{k}}$ are Fourier modes of dynamical variables. 
The coupling constant of the quadratic nonlinearity is denoted by $\mathcal{M}_{\sub{k}}^{\bm{p},\bm{q}}$. 
Linear operators, which describe a linear advection and/or forcing and/or dissipation, are represented by $\mathcal{L}_{\sub{k}}$. 
(Note that this formulation is easily extended to a general function beyond the Fourier modes if the coupling constant is defined appropriately.) 
One often finds a linear relation $Y_{\sub{k}}\! = \! \mathcal{K}_{\sub{k}}X_{\sub{k}}$ with an operator $\mathcal{K}_{\sub{k}}$ 
[eg. Eqs. (7) and (8) for $\dfgk$ and $\dpsik$] so that the above dynamical equation is rewritten as 
%
\begin{eqnarray}
\lbr \pdif{}{t} - \mathcal{L}_{\sub{k}} \rbr X_{\sub{k}} \e\  + \ \e \sum_{\bm{p}} \sum_{\bm{q}} \mathcal{N}_{\bm{k}}^{\bm{p},\bm{q}} 
X_{\sub{p}}^{\ast}X_{\sub{q}}^{\ast}\delta_{\bm{k}+\bm{p}+\bm{q},0} = 0\ , \\
\mathcal{N}_{\bm{k}}^{\bm{p},\bm{q}} \e \ \equiv\  \e \mathcal{M}_{\bm{k}}^{\bm{p},\bm{q}}\mathcal{K}_{\sub{p}}^{\ast}\ . 
\end{eqnarray}
%
It should be emphasized that since the term $X_{\sub{p}}^{\ast}X_{\sub{q}}^{\ast}\delta_{\bm{k}+\bm{p}+\bm{q},0}$ in Eq. (A.2) 
is symmetric for the interchange of $\bm{p}$ and $\bm{q}$, 
the anti-symmetric part of the coupling constant $\mathcal{N}_{\sub{k}}^{\bm{p},\bm{q}}$ never contribute to the dynamical equation. 

A balance equation for the squared norm $|X_{\sub{k}}|^{2}$ is given by
%
\begin{equation}
\pdif{}{t}|X_{\sub{k}}|^{2} - \mr{Re}X_{\sub{k}}^{\ast}\mathcal{L}_{\sub{k}}X_{\sub{k}} 
 + \sum_{\bm{p}} \sum_{\bm{q}} \mathcal{T}[\bm{k}|\bm{p},\bm{q}] \delta_{\bm{k}+\bm{p}+\bm{q},0} = 0\ , 
\end{equation}
%
where the triad transfer function, which gives finite contributions to the dynamical equation, is written as
%
\begin{eqnarray}
\mathcal{T}[\bm{k}|\bm{p},\bm{q}] \e \ = \ \e \mr{Re}\frac{1}{2} \lbr \mathcal{N}_{\sub{k}}^{\bm{p},\bm{q}} + \mathcal{N}_{\sub{k}}^{\bm{q},\bm{p}} \rbr 
X_{\sub{p}}^{\ast}X_{\sub{q}}^{\ast}X_{\sub{k}}^{\ast} \nonumber \\
\e \ = \ \e \mr{Re}\frac{1}{2} \lbr \mathcal{M}_{\sub{k}}^{\bm{p},\bm{q}}\mathcal{K}_{\sub{p}}^{\ast} + \mathcal{M}_{\sub{k}}^{\bm{q},\bm{p}}\mathcal{K}_{\sub{q}}^{\ast} \rbr 
X_{\sub{p}}^{\ast}X_{\sub{q}}^{\ast}X_{\sub{k}}^{\ast}\ .
\end{eqnarray}
%
Note that the above triad transfer function is symmetric for the interchange of $\bm{p}$ and $\bm{q}$, i.e., 
$\mathcal{T}[\bm{k}|\bm{p},\bm{q}] = \mathcal{T}[\bm{k}|\bm{q},\bm{p}]$. 
The equation (A.5) gives us a general definition of the symmetrized triad transfer function for a general dynamical equation 
with a quadratic nonlinearity. 

For the incompressible Navier-Stokes equation\cite{16}, the triad ``energy'' transfer function is derived, 
where the coupling constant is given by 
$\mathcal{M}_{\sub{k}}^{\bm{p},\bm{q}} \! = \mathcal{M}_{\sub{k}}^{\bm{q},\bm{p}} \! = \! \mathcal{P}_{lmn}[\bm{k}] 
\! = \! (i/2)[k_{m}(\delta_{ln}-k_{l}k_{n}/k^{2}) - k_{n}(\delta_{lm}-k_{l}k_{m}/k^2)]\ (l,m,n \! = \! 1,2,3)$, 
and $\mathcal{K}_{\sub{k}} \! = \! id.$ so that $X \! = \! Y \! = \! u_{l}\ (l \! = \! 1,2,3)$.
By making use of the incompressiblity condition, i.e., $\bm{k} \D \bm{u}_{\sub{k}}=0$, one obtains a detailed balance relation as follows, 
%
\begin{equation}
\mathcal{T}[\bm{k}|\bm{p},\bm{q}] + \mathcal{T}[\bm{p}|\bm{q},\bm{k}] + \mathcal{T}[\bm{q}|\bm{k},\bm{p}] = 0\ .
\end{equation}
%
    
For the gyrokinetic equation considered here, the the triad ``entropy'' transfer function, which is the same as in Eq. (17), is derived, 
where $\mathcal{M}_{\sub{k}}^{\bm{p},\bm{q}} \! = - \mathcal{M}_{\sub{k}}^{\bm{q},\bm{p}} \! = \! 
- (c/B)\bm{b} \! \D  ( \bm{p} \times \bm{q})$, $X \! = \! h_{\mr{s}}$, and $\mathcal{K}X \! = \! \dpsi$. 
(Note that $\mathcal{K}$ is a linear operator resulting from the quasi-neutrality, and the integral operators with a factor of $F_{\mr{Ms}}^{-1}$ 
is omitted in this description.) 
Then the detailed balance relation shown in Eq. (20) [also in Eq. (A.6)] straightforwardly obtained. 
The similar derivation is also applied to the Hasegawa-Mima equation\cite{35}, 
where $\mathcal{M}_{\sub{k}}^{\bm{p},\bm{q}} \! = - (c/B)\bm{b}\D( \bm{p} \times \bm{q}), \ X \! =\! \dphi,\ \mathcal{K}=-k^2$.

It should be noted that one finds the same gross value of 
$\sum_{\bm{p}}\sum_{\bm{q}}\mathcal{T}[\bm{k}|\bm{p},\bm{q}]\delta_{\bm{k}+\bm{p}+\bm{q},0}$ 
even if $\mathcal{T}[\bm{k}|\bm{p},\bm{q}]$ shown in Eq. (A.5) is replaced by the non-symmetrized one 
$\tilde{\mathcal{T}}[\bm{k},\bm{p},\bm{q}] \! \equiv \! \mr{Re}\mathcal{N}_{\sub{k}}^{\bm{p},\bm{q}}X_{\sub{p}}^{\ast}X_{\sub{q}}^{\ast}X_{\sub{k}}^{\ast}$, 
which has been used in earlier works\cite{27,28,29,30}. 
However, the magnitude of $\tilde{\mathcal{T}}[\bm{k},\bm{p},\bm{q}]$ itself involves a spurious value 
due to the contribution of anti-symmetric part of $\mathcal{N}_{\sub{k}}^{\bm{p},\bm{q}}$, which never contribute to the dynamical equation. 
For the quantitative evaluation of the detailed (or individual) entropy transfer processes, the symmetrized triad transfer function presented here must be used. 

%


\end{document}